\documentclass[superscriptaddress,twocolumn]{revtex4}
\usepackage{amsmath,lscape,epsfig}

\def\ii{\'{\i}}
\def\beq{\begin{equation}}
\def\eeq{\end{equation}}
\def\beqa{\begin{eqnarray}}
\def\eeqa{\end{eqnarray}}
\def\ban{\begin{eqnarray*}}
\def\ean{\end{eqnarray*}}
\def\bi{\begin{itemize}}
\def\ei{\end{itemize}}

\def\d{\mbox{d}}

\begin{document}

\title{The pasta phase within density dependent hadronic models}

\author{S. S. Avancini}
\affiliation{Depto de F\'{\i}sica - CFM - Universidade Federal de Santa
Catarina  Florian\'opolis - SC - CP. 476 - CEP 88.040 - 900 - Brazil}
\author{L. Brito}
\affiliation{Centro de F\ii sica Computacional - Department of Physics -
University of Coimbra - P-3004 - 516 - Coimbra - Portugal}
\author{J.R.Marinelli}
\affiliation{Depto de F\'{\i}sica - CFM - Universidade Federal de Santa
Catarina  Florian\'opolis - SC - CP. 476 - CEP 88.040 - 900 - Brazil}
\author{D.P.Menezes}
\affiliation{Depto de F\'{\i}sica - CFM - Universidade Federal de Santa
Catarina  Florian\'opolis - SC - CP. 476 - CEP 88.040 - 900 - Brazil}
\author{M.M.W. de Moraes}
\affiliation{Depto de F\'{\i}sica - CFM - Universidade Federal de Santa
Catarina  Florian\'opolis - SC - CP. 476 - CEP 88.040 - 900 - Brazil}
\author{C. Provid\^encia}
\affiliation{Centro de F\ii sica Computacional - Department of Physics -
University of Coimbra - P-3004 - 516 - Coimbra - Portugal}
\author{A.M.Santos}
\affiliation{Centro de F\ii sica Computacional - Department of Physics -
University of Coimbra - P-3004 - 516 - Coimbra - Portugal}

\begin{abstract}
In the present paper we investigate the onset of the pasta phase
with different parametrisations of the density dependent hadronic model 
and compare the results with one of the usual parametrisation of the
non-linear Walecka model. The influence of the scalar-isovector virtual
$\delta$ meson is shown. At zero
temperature two different methods are used, one based on coexistent phases
and the other on the Thomas-Fermi approximation. At finite temperature only
the coexistence phases method is used. $npe$ matter with fixed proton
fractions and in $\beta$-equilibrium are studied. We compare our results 
with  restrictions imposed on the the values of the density and pressure
at the inner edge of the crust, obtained from observations of the Vela pulsar 
and   recent isospin diffusion data from heavy-ion reactions, and with predictions from 
spinodal calculations. 
\end{abstract}
\maketitle

\vspace{0.50cm}
PACS number(s): {21.65.+f, 24.10.Jv, 26.60.+c, 95.30.Tg}
\vspace{0.50cm}

\section{Introduction}

Frustration is a phenomenon characterized by the existence of more than one 
low-energy configuration. The pasta phase is a frustrated system 
\cite{pethick,horo,maruyama}.
At densities of the order of 0.006 - 0.1 fm$^{-3}$ \cite{pasta1}
in neutral nuclear matter and 0.04 - 0.065 fm$^{-3}$ \cite{bao} in $\beta$-
equilibrium stellar matter, a competition between the strong and the
electromagnetic interactions takes place leading to a frustrated system.
The basic shapes of these complex structures were first named \cite{pethick} 
after well known types of cheese and pasta: 
droplets (bubbles = Swiss cheese), rods = spaghetti (tubes = penne) and 
slabs (lasagna) for three, two and one dimensions respectively. A 
droplet (bubble) and a rod (tube) have densities 
larger (smaller) than their surroundings, and are normally defined within a 
Wigner-Seitz cell. The pasta phase is the ground state configuration if its 
free energy per particle is lower than the corresponding to the homogeneous phase at the same 
density. The pasta phase is expected to exist somewhere between a solid and a
liquid phase, more like a liquid crystal \cite{pp98}. Its mechanical and thermal
properties are likely to depend on its shape and this study still 
remains to be done.

These pasta shapes at sub-nuclear densities  are expected to exist both in the 
crust of neutron stars (zero temperature, very low proton fraction, matter 
in $\beta$- equilibrium) and in supernova (finite temperature, proton fraction
around 0.3). In neutron stars the pasta phase coexists with a neutron gas; in 
supernova there is no neutron gas or it is very low in density 
\cite{watanabe05}. 

In a recent work \cite{pasta1} we have studied the 
existence of the pasta phase at zero and finite temperature within 
three different parametrisations of the
relativistic non-linear Walecka model (NLWM) \cite{sw}, namely 
NL3 \cite{nl3}, TM1 \cite{tm1} and GM3 \cite{glen}, the last one generally
used in the studies of stellar matter.
At zero temperature two different methods were used: the coexisting phases 
(CP) and the Thomas-Fermi (TF) approximation. We have checked that 
while the final equation of state (EoS) obtained with the 
different methods do not vary much, the internal structure varies
considerably.  The TF approximation was performed to test the much simpler CP
calculation and we have seen that the success of the CP calculation depends 
on the parametrisation of the surface energy for very small proton fractions
and close to the transition densities. At finite temperature only the CP 
method was used and compared with predictions from spinodal calculations. The 
pasta phase shrinks with the increase of the temperature and we have found that 
homogeneous matter can be the preferential phase also at very low 
densities depending on the temperature and the proton fraction. If $\beta$-
equilibrium is imposed the pasta phase does not appear in a CP
calculation. This indicates the necessity to use a good parametrisation for the
surface energy which is temperature, proton fraction and geometry dependent, 
as also stressed in \cite{gogelein,dalen}.

The authors of \cite{link99} have related the fraction of the moment of
  inertia contained in the crust of the Vela pulsar with the mass and the radius of the neutron
  star and the pressure and density at the crust-core interface. From
  realistic EoS they have obtained an expected range of values for the
  pressure at the inner edge of the crust and therefore also a relation
  between the radius and mass of the pulsar. This work shows the importance of 
understanding the exact density limits of the pasta phase
 and its consequences on the choice of appropriate equations of state.
More recently a new radius-mass relation for the Vela pulsar was obtained
taking as constraints recent 
isospin diffusion data from heavy-ion reactions \cite{bao}.
 In this work both the 
thermodynamical and the more accurate and reliable dynamical method were 
utilized in order to constrain the densities and related pressures of the 
pasta phase present in the crust of neutron stars. 
The pressures were obtained from the 
equation of state for neutron-rich nuclear matter constrained by isospin
diffusion data obtained in the same sub-saturation 
density range as the existing ones in the neutron star crust \cite{bao2}.

It is, however not clear how good are the predictions for the transition density
obtained from spinodal calculations. Clusterisation of the crust may have
been formed through equilibrium processes and it is important to compare
spinodal results with equilibrium results, obtained from the minimization of
the free energy. In \cite{pasta1} a first comparison was done and it was shown
that as a rule the transition densities obtained within an equilibrium
calculation are larger than the ones determined from the dynamical spinodals.

In the present work we use the same approximations (CP and TF) used in 
\cite{pasta1} to obtain the
pasta structures, but improve on the choice of the relativistic models, i.e.,
we obtain results with various density dependent hadronic models and 
investigate the influence of the delta mesons. 
We next justify our choices.

Density dependent hadronic models \cite{tw,gaitanos} have shown to provide 
richer and different results in many cases as compared with the simpler NLWM 
parametrisations \cite{inst04,inst062,peles}. In many situations the results 
are similar to the ones obtained with non-relativistic Skyrme-type models 
\cite{peles,bcmp07,del08,camille08}.

The inclusion of the isovector-scalar virtual $\delta (a_0(980))$ meson in 
hadronic effective field theories \cite{kubis97,liu} influences the 
calculation of the effective masses with 
important consequences on the symmetry energy, spinodals \cite{inst04}
and other quantities possibly related with the appearance of the pasta phase.
The $\delta$ field introduces  in the isovector channel the structure already existing 
in the isoscalar channel,  i.e., a balance between a scalar 
(attractive) and a vector (repulsive) potential.

In the following we consider three density dependent coupling
parametrisations,
TW \cite{tw}, DDH$\delta$ \cite{gaitanos} and GDFM \cite{gogelein} 
and two models with constant couplings, NL3 and NL3$\delta$.    
Neither NL3 nor TW include the $\delta$ meson.
A comparison is done between 
the transition pressures and densities from the
pasta phase to homogeneous matter obtained within the above mentioned models,
with the predictions obtained in  \cite{bao}.

The paper is organized as follows: in section II we briefly review the 
formalism underlying the homogeneous neutral $npe$ matter. 
In section III the pasta phase is built with the help of the coexisting phases
method and in section IV with the Thomas-Fermi approximation. In section  V
our results are displayed and commented and in section VI our conclusions are
drawn.

\section{The Formalism}

We consider a system of protons and neutrons with mass $M$
interacting with and through an isoscalar-scalar field $\phi$ with mass
$m_s$,  a isoscalar-vector field $V^{\mu}$ with mass
$m_\omega$, an isovector-vector field  $\mathbf b^{\mu}$ with mass
$m_\rho$ and an isovector-scalar field $\boldsymbol{\delta}$ with mass 
$m_\delta$. We also include a system of electrons with mass $m_e$. Protons and
electrons
interact through the electromagnetic field $A^{\mu}$.
The Lagrangian density reads:

\begin{equation}
\mathcal{L}=\sum_{i=p,n}\mathcal{L}_{i}+\mathcal{L}_e\mathcal{\,+L}_{{\sigma }}%
\mathcal{+L}_{{\omega }}\mathcal{+L}_{{\rho }}\mathcal{+L}_{{\delta }}\mathcal{+L}_{{\gamma }},
\label{lagdelta}
\end{equation}
where the nucleon Lagrangian reads
\begin{equation}
\mathcal{L}_{i}=\bar{\psi}_{i}\left[ \gamma _{\mu }iD^{\mu }-M^{*}\right]
\psi _{i}  \label{lagnucl},
\end{equation}
with 
\begin{eqnarray}
iD^{\mu } &=&i\partial ^{\mu }-\Gamma_{v}V^{\mu }-\frac{\Gamma_{\rho }}{2}{\boldsymbol{\tau}}%
\cdot \mathbf{b}^{\mu } - e \frac{1+\tau_3}{2}A^{\mu}, \label{Dmu} \\
M^{*} &=&M-\Gamma_{s}\phi-\Gamma_{\delta }{\boldsymbol{\tau}}\cdot \boldsymbol{\delta},
\label{Mstar}
\end{eqnarray}
and the electron  Lagrangian is given by
\begin{equation}
\mathcal{L}_e=\bar \psi_e\left[\gamma_\mu\left(i\partial^{\mu} + e A^{\mu}\right)
-m_e\right]\psi_e.
\label{lage}
\end{equation}
The meson and electromagnetic Lagrangian densities are 
\begin{eqnarray*}
\mathcal{L}_{{\sigma }} &=&\frac{1}{2}\left( \partial _{\mu }\phi \partial %
^{\mu }\phi -m_{s}^{2}\phi ^{2}\right)  \\
\mathcal{L}_{{\omega }} &=&\frac{1}{2} \left(-\frac{1}{2} \Omega _{\mu \nu }
\Omega ^{\mu \nu }+ m_{v}^{2}V_{\mu }V^{\mu } \right) \\
\mathcal{L}_{{\rho }} &=&\frac{1}{2} \left(-\frac{1}{2}
\mathbf{B}_{\mu \nu }\cdot \mathbf{B}^{\mu
\nu }+ m_{\rho }^{2}\mathbf{b}_{\mu }\cdot \mathbf{b}^{\mu } \right)\\
\mathcal{L}_{ {\delta }} &=&\frac{1}{2}(\partial _{\mu }\boldsymbol{\delta}%
\partial ^{\mu }\boldsymbol{\delta}-m_{\delta }^{2}{\boldsymbol{\delta}}^{2})\\ 
\mathcal{L}_{{\gamma }} &=&-\frac{1}{4}F _{\mu \nu }F^{\mu
  \nu },  
\end{eqnarray*}
where $\Omega _{\mu \nu }=\partial _{\mu }V_{\nu }-\partial _{\nu }V_{\mu }$
, $\mathbf{B}_{\mu \nu }=\partial _{\mu }\mathbf{b}_{\nu }-\partial _{\nu }\mathbf{b}%
_{\mu }-\Gamma_{\rho }(\mathbf{b}_{\mu }\times \mathbf{b}_{\nu })$ 
and $F_{\mu \nu }=\partial _{\mu }A_{\nu }-\partial _{\nu }A_{\mu }$.
The  parameters of the models  are: the nucleon mass $M=939$ MeV,
four density dependent coupling parameters $\Gamma_s$, $\Gamma_\omega$, $\Gamma_{\rho}$ and $\Gamma_{\delta}$ of the mesons to
the nucleons,  the electron mass $m_e$ and the electromagnetic coupling constant
$e=\sqrt{4 \pi/137}$.
In the above Lagrangian density $\boldsymbol {\tau}$ is
the isospin operator. 

\noindent From de Euler-Lagrange formalism we obtain coupled differential 
equations for
the scalar, vector, isovector-scalar, isovector-vector, electromagnetic and
nucleon fields. In the static case there are no currents and the spatial
vector components are zero. In \cite{pasta1} a complete description of the 
Thomas-Fermi approximation applied to different parametrisations of the NLWM 
is given. As the differences arising from the inclusion of the $\delta$- mesons
and the use of density dependent couplings are small, we do not 
repeat the equations here. A rearrangement term is the landmark of most 
density dependent hadronic models \cite{fuchs,br} and the simple mean field 
approximation (MFA) is outlined next so that its appearance is better 
understood. The equations of motion for the  fields can be obtained
and solved self-consistently in the MFA (the photon and
meson fields are classical fields), neglecting states of negative energy 
(no-sea approximation)\cite{tw}. 

The meson fields within the mean field approximation are obtained from the
following equations:
\begin{equation}
  m_s^2\phi_0= \Gamma_s \rho_s,
\label{elphi}
\end{equation}
\begin{equation}
 m_\omega^2 V_0 = \Gamma_\omega \rho,
\label{elV0}
\end{equation}
\begin{equation}
m_\rho^2 b_0 =\frac{\Gamma_\rho}{2} \rho_3,
\label{elb0}
\end{equation}
\begin{equation}
m_\delta^2 \delta_3 = \Gamma_\delta \rho_{s3}.
\label{eld0}
\end{equation}
The second members of the above equations include the 
the  equilibrium  densities $\rho =$ $\rho _{p}+\rho _{n}$,
$\rho_3 =\rho _{p}-\rho_{n} $, $\rho_{s} =$ $\rho _{sp}+\rho_{sn} $ and $\rho_{s3} =$ $\rho _{sp}-\rho_{sn} $
where the proton/neutron densities  are given by
\begin{equation}
\rho_i=\frac{1}{\pi^2} \int {p^2 dp}(f_{i+}-f_{i-}),\,\, i=p,n
\end{equation}
and the corresponding scalar density by
\begin{equation}
\rho _{s_{i}}=\frac{1}{\pi^2} \int {p^2 dp}
\frac{M_{i}^{*}}{\sqrt{p^{2}+{M_{i}^{*}}^{2}}}(f_{i+}+f_{i-}),
\label{rhoscalar}
\end{equation}
with the distribution functions given by
\begin{equation}
f_{i \pm}=\frac{1}{1+\exp[(\epsilon_i^{\ast}({\mathbf p}) \mp \nu_i)/T]}\;,
\label{distf}
\end{equation}

where
${\epsilon}_i^{\ast}=\sqrt{{\mathbf p}^2+{M_i^*}^2}$,
\begin{equation}
M_i^*=M - \Gamma_s ~ \phi_0 - \tau_{3 i}~ \Gamma_{\delta}~ \delta_3,
\label{effm}
\end{equation}
and the effective chemical potentials are
\begin{equation}
\nu_i=\mu_i - \Gamma_\omega V_0 - \frac{\Gamma_{\rho}}{2}~  \tau_{3 i}~ b_0-{\Sigma^{R}_0} ,
\end{equation}
$\tau_{3 i}= \pm 1$ is the isospin projection for the protons and neutrons
respectively.
The density dependent models in the mean field approximation contain a rearrangement term ${\Sigma^{R}_0}$ \cite{gaitanos}:

\begin{equation*}
{\Sigma^{R}_{0}}=
\frac{\partial \Gamma_{v}}{\partial \rho} \rho V_0 +
\frac{\partial \Gamma_{\rho}}{\partial \rho} \rho_3 ~ \frac{b_0}{2} -
\frac{\partial \Gamma_s}{\partial \rho} \rho_{s} \phi_0
-\frac{\partial\, \Gamma_\delta}{\partial \rho}\, \rho_{s3}\, \delta_3.
\end{equation*}
Notice that for $T=0$ MeV the distribution function for particles given in
equation (\ref{distf}) becomes the simple step function
$f_{i}=\theta(P_{Fi}^2-p^2)$ and the distribution function for
anti-particles vanishes.

In the description of the equations of state
of a system, the required quantities are the baryonic density,  energy
density,  pressure and free energy. The energy density reads:

\begin{equation}
{\cal E}=\sum_{i=n,p} K_{i}+{\cal E}_{\sigma }+{\cal E}_{\omega }+{\cal E}_{\delta }+{\cal E}_{\rho},
\label{EQ:E}
\end{equation}
with
\begin{eqnarray}
K_{i} &=& \frac{1}{\pi^2} \int p^2 dp {\sqrt{p^{2}+{
M_{i}^{*}}^{2}}} \left( f_{i+}+f_{i-}\right), \label{EQ:EK} \\
{\cal E}_{\sigma } &=&\frac{m_{s}^{2}}{2}\phi _{0}^{2},  \label{EQ:Esigma} \\
{\cal E}_{\omega } &=&\frac{m_{v}^{2}}{2}V_{0}^{2},
\label{EQ:Eomega}
\\
{\cal E}_{\delta } &=&\frac{m_{\delta }^{2}}{2}\delta _{3}^{2},  \label{EQ:Edelta} \\
{\cal E}_{\rho } &=&\frac{m_{\rho }^{2}}{2}b_{0}^{2}.
\label{EQ:Erho}
\end{eqnarray}

The pressure is given by:
\begin{equation}
P=\sum_{i=n,p}P_{i}+P_{\sigma }+P_{\omega }+P_{\delta }+P_{\rho },
\label{pressnlwm}
\end{equation}
with the partial pressures associated with the nucleons and the various fields
\begin{eqnarray*}
P_{i} &=&\frac{1}{3\pi ^{2}}\int dp\frac{p^{4}}{\sqrt{p^{2}+{M_i^{*}}^{2}}}%
\left( f_{i+}+f_{i-}\right),  \\
P_{\sigma } &=& -\frac{m_{s}^{2}}{2}\phi _{0}^{2}\left(1+%
2\frac{\rho}{\Gamma_s}\frac{\partial \Gamma_s}{\partial \rho}\right), \\
P_{\omega } &=&\frac{m_{v}^{2}}{2}V_{0}^{2}\left(1+%
2\frac{\rho}{\Gamma_\omega}\frac{\partial \Gamma_\omega}{\partial \rho}\right),\\
P_{\rho } &=&\frac{m_{\rho }^{2}}{2}b_{0}^{2}\left(1+%
2\frac{\rho}{\Gamma_\rho}\frac{\partial \Gamma_\rho}{\partial \rho}\right),\\
P_{\delta } &=&-\frac{m_{\delta }^{2}}{2}\delta _{3}^{2}\left(1+%
2\frac{\rho}{\Gamma_\delta}\frac{\partial \Gamma_\delta}{\partial \rho}\right).
\end{eqnarray*}

\noindent The free energy density is defined as:
\begin{equation}
{\cal F}= {\cal E}-T {\cal S},
\end{equation}

\noindent with the entropy density :
\begin{equation}
{\cal S}= \frac{1}{T}({\cal E}+P-\mu_p \rho_p - \mu_n \rho_n).
 \end{equation}

As for the electrons, their density and distribution functions read:
\begin{equation}
\rho_e=\frac{1}{\pi^2} \int {p^2 dp}(f_{e+}-f_{e-}),
\label{rhoe}
\end{equation}
with
\begin{equation}
f_{e\pm}({\mathbf r},{\mathbf p},t)\,=\,\frac{1}{1+\exp[(\epsilon_e\mp
\mu_e)/T]},
\end{equation}
where $\mu_e$ is the electron chemical potential and
$\epsilon_e=\sqrt{p^2+m_e^2}$. 
We always consider neutral matter and therefore the electron density is
equal to the proton density. In the calculation of the non-homogeneous phase
the Coulomb energy of the proton and electron distributions
is included. We study both matters (homogeneous and pasta structured), 
with a fixed proton fraction, as we get in heavy ion collisions (although in 
this case matter is not neutral) and in $\beta$-equilibrium as in stellar  matter. In the latter,
charge neutrality conditions fix the electron chemical potential and density.
The onset of muons occurs
above the transition density to homogeneous phase and therefore  
the proton density remains equal to the electron density.

The energy density, pressure, free energy density and entropy density of the 
electrons are
\begin{equation}
{\cal E}_e=\frac{1}{\pi^2} 
\int \d p\, p^2  \sqrt{p^2+m_e^2} \left(f_{e+}+f_{e-}\right),
\end{equation}

\begin{equation}
P_e=\frac{1}{3 \pi^2} 
\int \d p\,\frac{p^4}{\sqrt{p^2+m_e^2}} \left(f_{e+}+f_{e-}\right),
\end{equation}

\begin{equation}
{\cal F}_e= {\cal E}_e-T {\cal S}_e,
\end{equation}

{\noindent and}

\begin{equation}
{\cal S}_e= \frac{1}{T}({\cal E}_e+P_e-\mu_e \rho_e ).
 \end{equation}

To obtain the equations for the TW parametrisation \cite{tw} of the 
density dependent hadronic model, all information on the $\delta$ meson is
excluded. 
For the  NL3 \cite{nl3} parametrisation the density dependent parameters are
substituted by the usual 
coupling constants $g_\sigma$, $g_\omega$, $g_\rho$  and
non-linear parameters are included (see \cite{pasta1}, for instance). 
The NL3$\delta$ parametrisation is defined with the same values for $g_\sigma$ and $g_\omega$  as 
in the NL3  parametrisation, $g_\rho =14.29$ and $g_\delta =7.85$, in 
such a way that the symmetry energy has the same value at the saturation
density as the 
NL3 parametrisation.
We show in  Table \ref{prop} the nuclear matter 
properties reproduced by the models we discuss in the present work.

\begin{table}[h]
\caption{ Nuclear matter properties at the saturation density.}
\label{prop}
\begin{center}
\begin{tabular}{cccccccccc}
\hline
&  NL3/NL3$\delta$ & TW   & DDH$\delta$ &GDFM \\
&   \cite{nl3} & \cite{tw} &  \cite{gaitanos} & \cite{gogelein}\\
\hline
\hline
$B/A$ (MeV) & 16.3  & 16.3  & 16.3&16.25\\
$\rho_0$ (fm$^{-3}$) & 0.148  & 0.153  & 0.153& 0.178\\
$K$ (MeV) & 272  & 240 & 240& 337 \\
${\cal E}_{sym.}$ (MeV)  & 37.4  & 32.0 & 25.1 & 32.11 \\
$M^*/M$ & 0.60 & 0.56  & 0.56& 0.68\\
$L$ (MeV) & 117/148 & 55 & 44 & 57\\
$T_c$ (MeV) & 14.55 & 15.18 & 15.18 & 15.95\\
\hline
\end{tabular}
\end{center}
\end{table}

The density-dependent coupling parameters are adjusted in order to reproduce
some of the nuclear matter bulk properties, using the
following parametrisation for the  $\sigma$ and $\omega$ mesons:
\begin{equation}
\Gamma_i(\rho)=\Gamma_i(\rho_0)h_i(x), \quad x=\rho/\rho_0
\label{paratw1}
\end{equation}
with
\begin{equation}
h_i(x)=a_i \frac{1+b_i(x+d_i)^2}{1+c_i(x+d_i)^2},  \quad  i=s,v,
\end{equation}
and 
\begin{equation}
h_{\rho}(x)=\exp[-a_{\rho}(x-1)]
\end{equation}
for the TW model \cite{tw}.
In the case of the DDH$\delta$ model \cite{gaitanos} we use the TW
parametrisation for $\Gamma_\sigma$ and $\Gamma_\omega$ and for the $\rho$ and $\delta$
mesons we take:

$$h_i(x)=a_i \exp[-b_i(x-1)]-c_i(x-d_i), \qquad i=\rho,\,\delta.$$
The parameters $a_i,b_i,c_i$ and $d_i$ are given in Table \ref{para}
and $\rho_0$ is the saturation density.

\begin{table}[h]
\caption{Parameters of the TW \cite{tw} and  DDH$\delta$ models \cite{gaitanos,inst04}}
\label{para}
\begin{center}
\begin{tabular}{lcccccc}
\hline
i&$m_i$(MeV)&$\Gamma_i(\rho_0)$&$a_i$&$b_i$&$c_i$&$d_i$\\
\hline
$\sigma$&550&10.72854&1.365469&0.226061&0.409704&0.901995\\
$\omega$&783&13.29015&1.402488&0.172577&0.344293&0.983955\\
$\rho_{\rm TW}$ &763& 7.32196&0.515&--- &--- &--- \\
$\rho_{{\rm DDH}\delta}$ &763&11.727&0.095268&2.171&0.05336&17.8431\\
$\delta$&980&7.58963&0.01984&3.4732 &-0.0908&-9.811\\
\hline
\end{tabular}
\end{center}
\end{table}

The density dependent parametrisation GDFM obtained
in \cite{gogelein} takes into account the renormalization of the
relativistic mean-field theory due to Fock exchange terms. 
It ensures a good description of the properties of the equation of state at
high density as obtained  
with calculations for asymmetric nuclear matter\cite{dalen} with 
Dirac-Brueckner-Hartree-Fock calculations.

The GDFM parametrisation for all four mesons coupling parameters reads:
\begin{equation}
\Gamma_i(\rho)=a_i + (b_i+d_{i}x^3) \exp(-c_{i}x).                      
\end{equation}
Around the saturation density  a correction to the coupling parameter for the $\omega$ meson is introduced:

\begin{equation}
\Gamma_{\omega {\rm cor}}(\rho)=\Gamma_{\omega}(\rho)-a_{\rm cor}
\exp\left[-\left({\frac{\rho-\rho_0}{b_{\rm cor}}}\right)^2 \right],
\end{equation}

{\noindent where $ a_{\rm cor}=0.014$ and $ b_{\rm cor}=0.035\, {\rm fm}^{-3}$. }

\begin{table}[h]
\caption{Parameters of the GDFM model \cite{gogelein}}
\label{paragog}
\begin{center}
\begin{tabular}{lcccccc}
\hline
i&$m_i$(MeV)&$a_i$&$b_i$&$c_i$&$d_i$\\
\hline
$\sigma$&550&7.7868&2.58637&2.32431&3.11504\\
$\omega$&782.6&9.73684&2.26377&7.05897&---\\
$\rho$&769&4.56919&5.45085&1.20926&---\\
$\delta$&983&2.68849&6.7193 &0.503759&0.403927\\
\hline
\end{tabular}
\end{center}
\end{table}

\begin{figure}
\begin{center}
\includegraphics[width=7.cm]{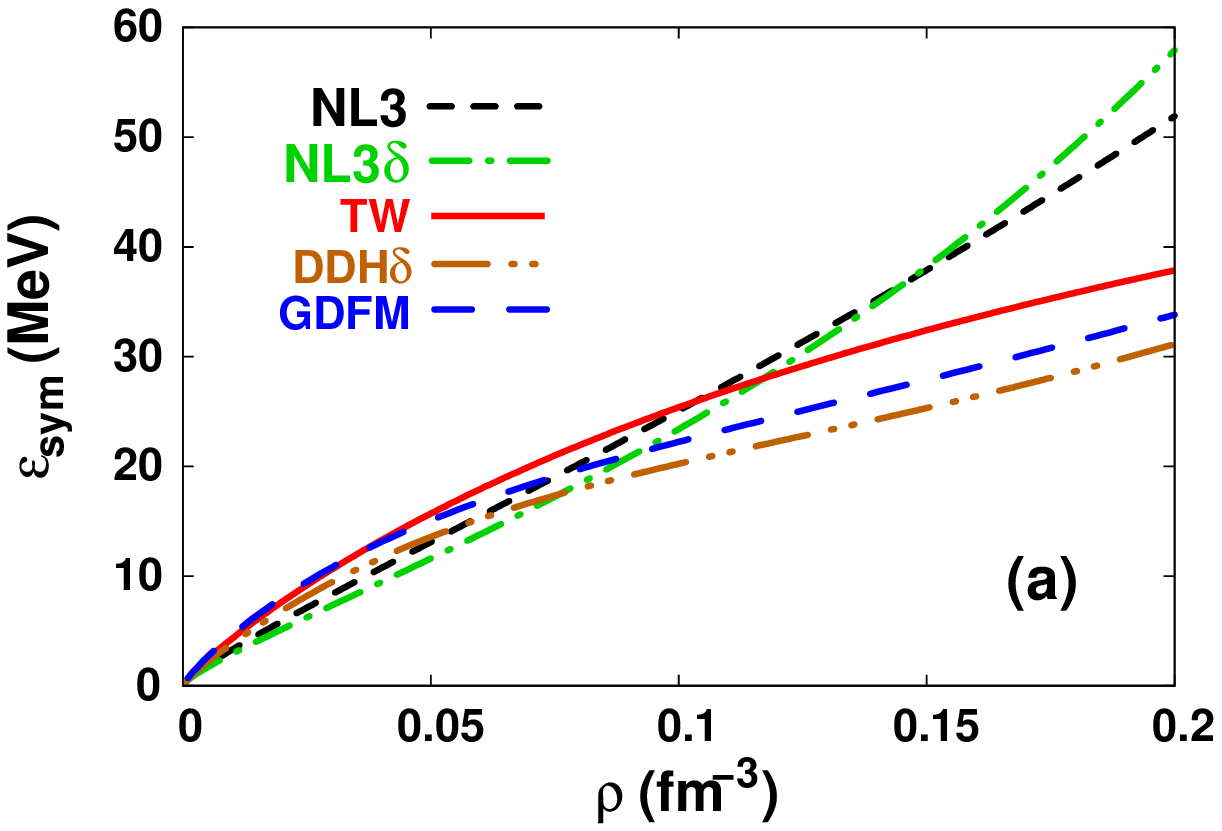} \\
\includegraphics[width=7.cm]{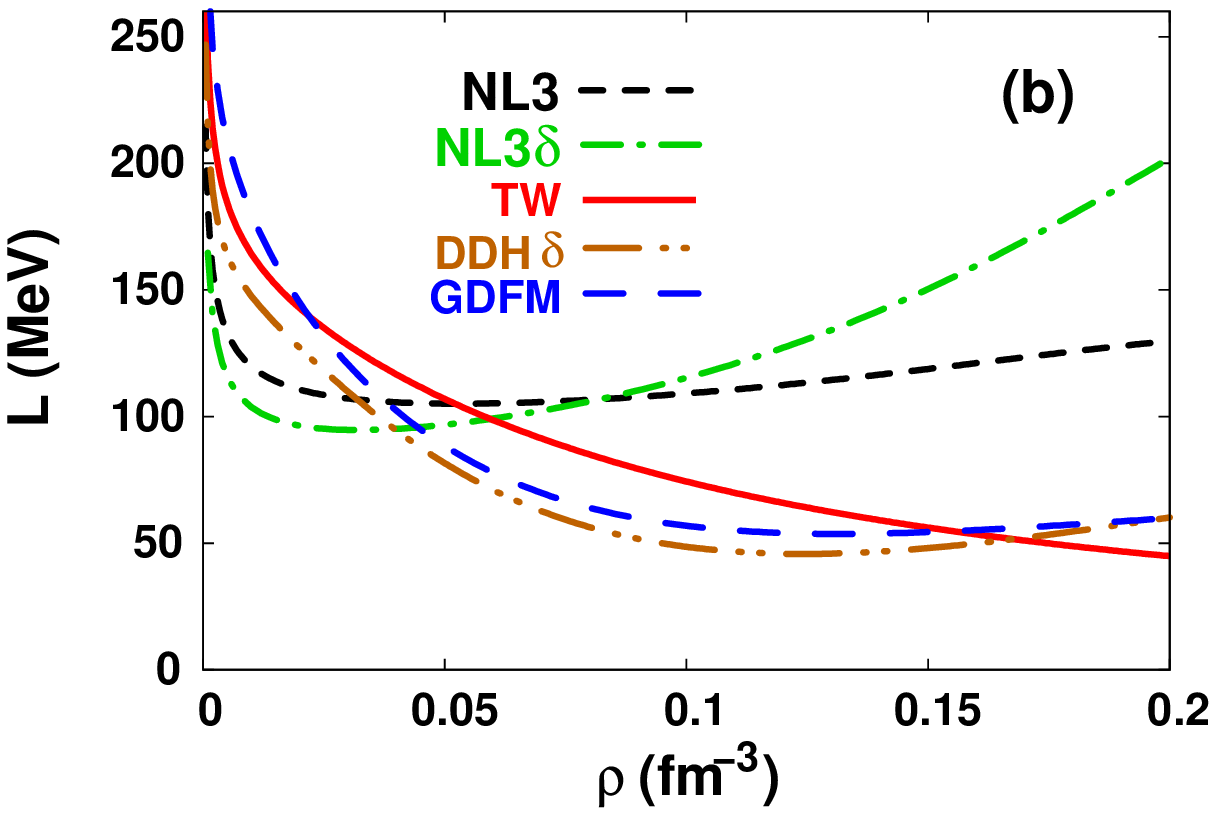} \\
\end{center}
\caption{a) Symmetry energy and b) slope of symmetry energy for the 
models under study.}
\label{fig1}
\end{figure}

The properties we discuss in the following depend on the isovector channel 
of the
nuclear force, mainly the results obtained for $\beta$-equilibrium matter.
Therefore we show in Fig. \ref{fig1} the symmetry energy and the slope  of the
symmetry energy $L= 3\, \rho_0\, \partial \epsilon_{sym}/\partial \rho$. This
quantity defined at the saturation density is given in Table \ref{prop}.
The symmetry energy of NLW models becomes quite hard for densities above
$\sim$ 0.1 fm$^{-3}$. However, at subsaturation densities, the $\delta$ meson
gives rise to a softer symmetry energies: this is true both for NL3$\delta$
and for DDH$\delta$ and GDFM,  DDH$\delta$ being softer than GDFM. Looking at
the slope of the symmetry energy we see that GDFM and DDH$\delta$ show a very
similar behavior. Except for TW all models show at low densities a decrease
of the slope followed by an increase of the slope above $\rho\sim0.05$
fm$^{-3}$ for NLW models or $\rho\sim0.12$ fm$^{-3}$ for density dependent 
models with the $\delta$ meson.
 For TW the slope always decreases with density  more slowly than all the
other  models.

\section{Coexisting phases}

\subsection{Nuclear pasta}

As in \cite{pasta1,maruyama}, for a given total density $\rho$ and proton 
fraction $Y_p=\rho_p/\rho$ the pasta structures are built with different 
geometrical forms in a background nucleon gas. This is achieved by calculating 
from the Gibbs' conditions the  
density and the proton fraction of the pasta and
of the background gas, so that in the whole we have to  
 solve simultaneously the following eight equations:
\begin{equation}
P^I(\nu_p^I,\nu_n^I,{M^*_n}^I,{M^*_p}^I)=P^{II}(\nu_p^{II},\nu_n^{II},
{M^*_n}^{II},{M^*_p}^{II}),\label{gibbs1}
\end{equation}
\begin{equation}
\mu_i^I=\mu_i^{II}, \quad i=p,n\label{gibbs2}
\end{equation}
\begin{equation}
m_\sigma^2 \phi_0^I = \Gamma_\sigma \rho_s^I,\label{gibbs3}
\end{equation}
\begin{equation}
m_\sigma^2 \phi_0^{II} = \Gamma_\sigma \rho_s^{II},\label{gibbs4}
\end{equation}
\begin{equation}
m_\delta^2 \delta_3^I = \Gamma_\delta \rho_{s3}^I, \label{gibbs5}
\end{equation}
\begin{equation}
m_\delta^2 \delta_3^{II} = \Gamma_\delta \rho_{s3}^{II}, \label{gibbs6}
\end{equation}
\begin{equation}
f \rho_p^I + (1-f) \rho_p^{II} = \rho_p = Y_p \rho,\label{gibbs7}
\end{equation}
where I and II label each of the phases, $f$ is the volume fraction of 
phase I: 
\begin{equation}
f= \frac{\rho -\rho^{II}}{\rho^I-\rho^{II}} 
\end{equation}
and $Y_p$ is the global proton fraction. The density of electrons is
uniform and taken as $\rho_e=Y_p \rho$. For the NL3 and NL3$\delta$ 
parametrizations, non-linear $\sigma$ terms must be included in (\ref{gibbs3}) and (\ref{gibbs4}).
 
The total pressure is given  by $P=P^I+P_e$. 
The total energy density of the system is given by
\begin{equation}
{\cal E}= f {\cal E}^I + (1-f) {\cal E}^{II} + {\cal E}_e +
{\cal E}_{surf} + {\cal E}_{Coul}, 
\label{totener}
\end{equation}
where, by minimizing the sum  ${\cal E}_{surf} + {\cal E}_{Coul}$ with respect
to the size of the droplet/bubble, rod/tube or slab we get 
\cite{maruyama}
${\cal E}_{surf} = 2 {\cal E}_{Coul},$ and  
\begin{equation}
{\cal E}_{Coul}=\frac{2 \alpha}{4^{2/3}}(e^2 \pi \Phi)^{1/3} 
\left(\sigma D (\rho_p^I-\rho_p^{II})\right)^{2/3},
\end{equation}
where $\alpha=f$ for droplets and $\alpha=1-f$ for bubbles, 
 $\sigma$ is the surface energy coefficient,
$D$ is the dimension of the system. For droplets, rods and slabs,
\begin{equation}
\Phi=\begin{cases}
\left(\frac{2-D f^{1-2/D}}{D-2}+f \right) \frac{1}{D+2}, \quad D=1,3;\\
 \frac{f-1-ln(f)}{D+2}, \quad D=2. \end{cases}
\end{equation}
and for bubbles and tubes the above expressions are valid with 
$f$ replaced by $1-f$.

Concerning the surface energy, the authors of \cite{maruyama} state that, in
this case, the appearance of the pasta phase essentially depends on the value
of the surface tension. We have fixed the surface tension at different values
and confirmed their claim. We have  
parameterized the surface energy coefficient in terms of the proton fraction
according to the functional proposed in  \cite{lattimer}, which was obtained
by fitting Thomas-Fermi and Hartree-Fock numerical values with a Skyrme force,
\begin{equation}
\sigma=\sigma_0 \frac{16+b}{\frac{1}{Y_p^3}+\frac{1}{(1-Y_p)^3}+b}h_t,
\label{sigma}
\end{equation}
with 
\begin{equation}
h_t=\left[1-\left(\frac{T}{4T_c Y_p(1-Y_p)}\right)^2\right]^2,
\end{equation}
$\sigma_0=1.03$ MeV/fm$^2$ and $b=24.4$ and $T_c$ is the critical temperature
above which there is a smooth transition from the gas phase to the liquid
phase \cite{inst062} and given in Table \ref{prop}. We have checked that small 
variations of this temperature do not affect our results. The proton fraction 
considered throughout the calculation of $\sigma$ is the one of the denser 
phase.

Each structure is considered to be in the center of a charge neutral 
Wigner-Seitz  cell
constituted by neutrons, protons and leptons \cite{shen}. 
The Wigner-Seitz cell
is a sphere/cilinder/slab whose volume is the same as the unit BCC cell. 
In \cite{shen} the
internal structures are associated with heavy nuclei. Hence, the radius of the
droplet (rod,slab) and of the Wigner-Seitz cell are
respectively given by:
\begin{equation}
R_D=\left( \frac{\sigma D}{4 \pi e^2 (\rho_p^I-\rho_p^{II})^2 \Phi} \right)^{1/3}
\quad R_W=\frac{R_D}{(1-f)^{1/D}}. 
\end{equation}
  
\subsection{Stellar pasta}

In this case, hadronic matter is in $\beta$ equilibrium.
The condition of $\beta$ equilibrium in a system of protons, neutrons,
electrons and muons is 
\begin{equation}
\mu_p=\mu_n-\mu_e,
\end{equation}
where $\mu_e =\mu_\mu$.  As the muons are added, the imposition of charge 
neutrality requires that 
\begin{equation}
\rho_p=\rho_e + \rho_\mu.
\end{equation}
The Gibbs conditions to be enforced are  
\begin{equation}
\mu_n^I=\mu_n^{II}, \quad \mu_e^I=\mu_e^{II},
\end{equation}
and
\begin{equation}
f \left(\rho_p^I-\rho_e^I-\rho_\mu^I\right) + (1-f) \left(\rho_p^{II} -
\rho_e^{II} - \rho_\mu^{II}\right)=0 \label{eqb} 
\end{equation}
together with  (\ref{gibbs1}), (\ref{gibbs3}), (\ref{gibbs4}), (\ref{gibbs5}) 
and (\ref{gibbs6}).
Here the density of electrons is no longer taken uniform as in the
  last section, but appears as the solution of equation (\ref{eqb}). 
The densities of interest to the study of the pasta phase
  are too low for the muons to appear, which generally occurs for
    densities above $0.1$ fm$^{-3}$ \cite{camille08}.

\section{Pasta-phase within the Thomas-Fermi approximation}

In the present work we repeat the same numerical prescription given in 
\cite{pasta1} where, within the Thomas-Fermi approximation of the 
non-uniform {\it npe} matter, the fields are assumed to vary slowly so that 
the baryons can be treated as moving in locally constant fields at each point. 
In the Thomas-Fermi approximation, the energy is a functional of the density 
given by:
\begin{equation}
E_{TF}=\int d^3 r \left( \sum_{i=p,n,e} E_i(\mathbf r) 
\right. \nonumber 
\end{equation}
\begin{equation}
+\frac{1}{2} [(\nabla \phi_0(\mathbf r))^2 +  
m_\sigma^2\phi_0^2(\mathbf r)]   
-\frac{1}{2}[(\nabla V_0(\mathbf r))^2 +m_\omega^2 V_0^2(\mathbf r)] 
\nonumber 
\end{equation}
\begin{equation}
 -\frac{1}{2} [(\nabla b_0(\mathbf r) )^2 
+m_{\rho}^2 b_0^2(\mathbf r)]
-\frac{1}{2} [(\nabla \delta_3(\mathbf r) )^2 
+m_{\delta}^2 \delta_3^2(\mathbf r)]
\end{equation}
\begin{equation}
+\Gamma_\omega V_0(\mathbf r)\rho + \frac{1}{2} \Gamma_\rho b_0(\mathbf r) \rho_3 
\end{equation}
\begin{equation}
\left.
-\frac{1}{2}[\nabla A_0(\mathbf r) ]^2 + e(\rho_p-\rho_e)A_0(\mathbf r)  \right) ~~,
\label{enfun} 
\end{equation}
where 
\begin{equation}
E_i=\frac{1}{\pi^2}\int_0^{p_{F_i}(\mathbf r)} dp\, p^2 (p^2+{M^{\star}}^2)^{1/2} ~~,i=p,n~~, 
\end{equation}
and 
\begin{equation}
E_e=\frac{1}{\pi^2}\int_0^{p_{F_e}(\mathbf r)} dp\, p^2 (p^2+m_e^2)^{1/2} ~~.
\end{equation}

The definition of the thermodynamic potential reduces to
\begin{equation}
\Omega=E_{TF}[\rho_i] -\sum_{i=n,p,e} \mu_i \int dr \rho_i(\mathbf r) ~.
\label{pote}
\end{equation}
The minimization of the  above functional with the constraint of a fixed number of protons, 
neutrons and electrons yields the equations:
$$(p^2_{F_p}(\mathbf r)+{M^\star_p}^2(\mathbf r) )^{1/2} +\Gamma_\omega V_0(\mathbf r) 
+ \frac{1}{2}\Gamma_\rho b_0(\mathbf r) + \Sigma^R_0 $$
\begin{equation}
+eA_0(\mathbf r)=\mu_p \, 
\end{equation} 
$$(p^2_{F_n}(\mathbf r)+{M^\star_n}^2(\mathbf r) )^{1/2} +\Gamma_\omega V_0(\mathbf r) 
- \frac{1}{2}\Gamma_\rho b_0(\mathbf r)  + \Sigma^R_0$$ 
\begin{equation}
=\mu_n, 
\end{equation}
where $M^\star_p$ and $M^\star_n$ are given in eq.(\ref{effm}) and
\begin{equation}
(p^2_{F_e}(\mathbf r)+m_e^2)^{1/2}-e A_0(\mathbf r)  =\mu_e.
\end{equation}

The numerical algorithm for the description of the neutral $npe$ matter
was discussed in detail in \cite{pasta1}.
The Poisson equation is always solved by using the appropriate Green 
function according to the  spatial dimension of interest and the
Klein-Gordon equations are solved by expanding the meson fields in a harmonic 
oscillator basis with one, two or three dimensions based on the method 
proposed in \cite{ring}.

\begin{table*}
\caption{Transition densities in fm$^{-3}$ and corresponding pressures (CP and TF calculations) for 
the non-homogeneous to homogeneous phase at the inner edge of the crust ($T=0$)MeV.}
\label{tzero}
\begin{center}
\begin{tabular}{cccccccc}
\hline
model & EoS & dynamical & thermodynamical & pasta (CP)  & P (CP)& pasta (TF) &
P (TF) \\
      &     & spinodal &  spinodal  &      versus &        MeV/fm$^3$    & versus & MeV/fm$^3$ \\
      &     & versus EoS & versus EoS & uniform matter &      & uniform matter & \\
\hline
NL3  & $y_p=0.5$       & 0.083  & 0.096  &  0.096  & 2.65   & 0.112 & 2.64    \\
NL3  & $y_p=0.3$       & 0.080  & 0.094  &  0.079  & 1.05  & 0.100 & 1.14   \\
NL3  & $\beta$-equil. & 0.053   & 0.065  & -   & -  &  0.054 & 0.24 \\
\hline
NL3$\delta$  & $y_p=0.5$       & 0.083  & 0.096  &  0.096  & 2.65  & 0.112 & 2.64    \\
NL3$\delta$  & $y_p=0.3$       & 0.079  & 0.093  &  0.080   &1.06   & 0.099 & 1.14    \\
NL3$\delta$  & $\beta$-equil. & 0.048  &  0.056  &  -  & -  &
0.051 & 0.16    \\
\hline
TW  & $y_p=0.5$       &0.083  & 0.096  &  0.098  & 2.74  & 
0.113 & 2.66   \\
TW  & $y_p=0.3$       &  0.084 &0.095  & 0.099   & 1.40  & 0.111 & 1.27    \\
TW  & $\beta$-equil. & 0.075  & 0.085  & 0.060   & 0.26  & 0.076 & 0.40   \\
\hline
DDH$\delta$  & $y_p=0.5$       &0.083  & 0.096  & 0.098   & 2.74   & 0.113 & 2.66    \\
DDH$\delta$  & $y_p=0.3$       & 0.084  & 0.094  &0.107    &1.55   & 0.115 & 1.29  \\
DDH$\delta$  & $\beta$-equil. & 0.079  & 0.085  &0.089  & 0.17
& 0.079 &  0.10   \\
\hline
GDFM  & $y_p=0.5$       & 0.133  & 0.141  &  0.119  & 3.55  & 0.144 & 3.81    \\
GDFM  & $y_p=0.3$       & 0.131  & 0.138  & 0.119   & 1.79  & 0.140 & 1.83    \\
GDFM  & $\beta$-equil.  & 0.051  & 0.058  & 0.027   &0.04   & 0.052 & 0.13    \\
\hline
\end{tabular}
\end{center}
\end{table*}

\section{Results and discussions}

We show and discuss the results obtained using the coexistence phases
(CP) and the Thomas-Fermi (TF) methods in the framework of the several 
relativistic models presented, always for $npe$ matter. 
We start with the results at $T=0$ MeV.

\begin{figure}[htb]
\begin{center}
\begin{tabular}{cc}
\includegraphics[width=4.5cm]{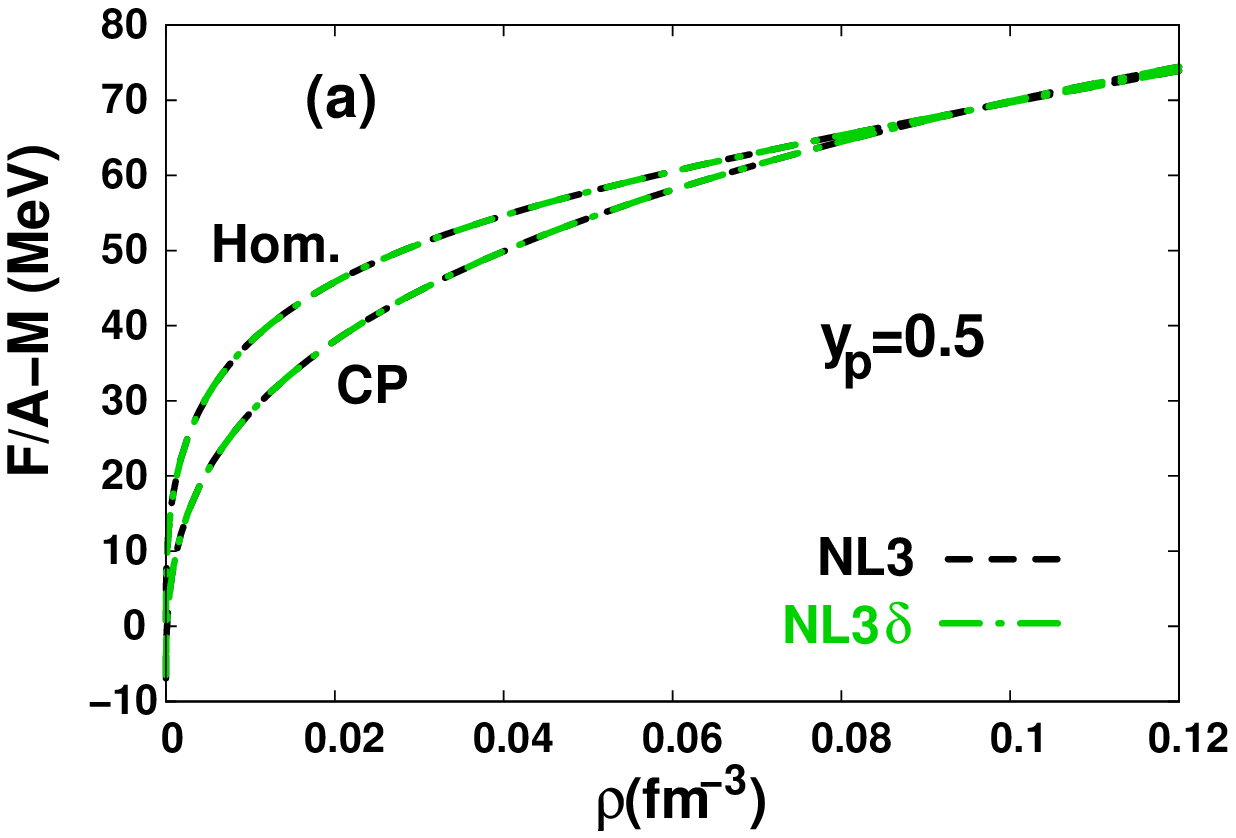} &
\includegraphics[width=4.5cm]{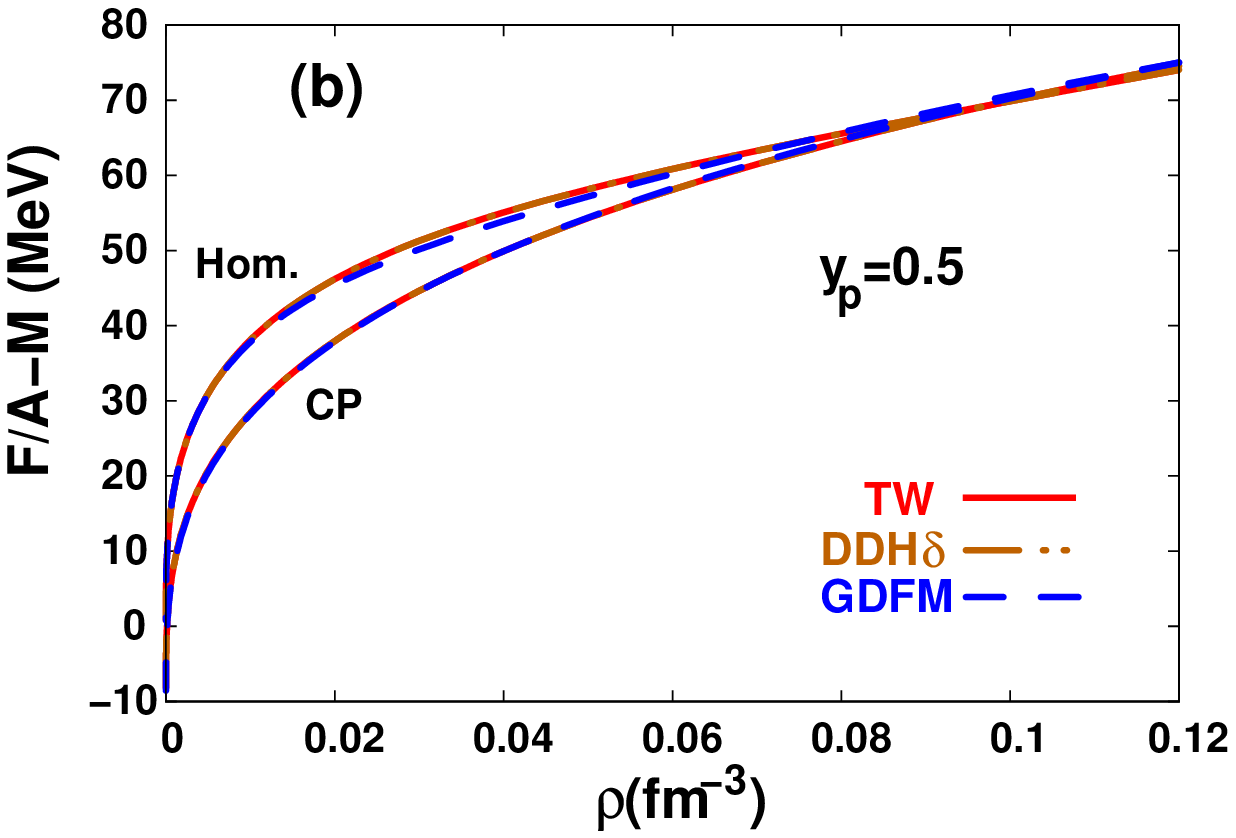}  \\
\includegraphics[width=4.5cm]{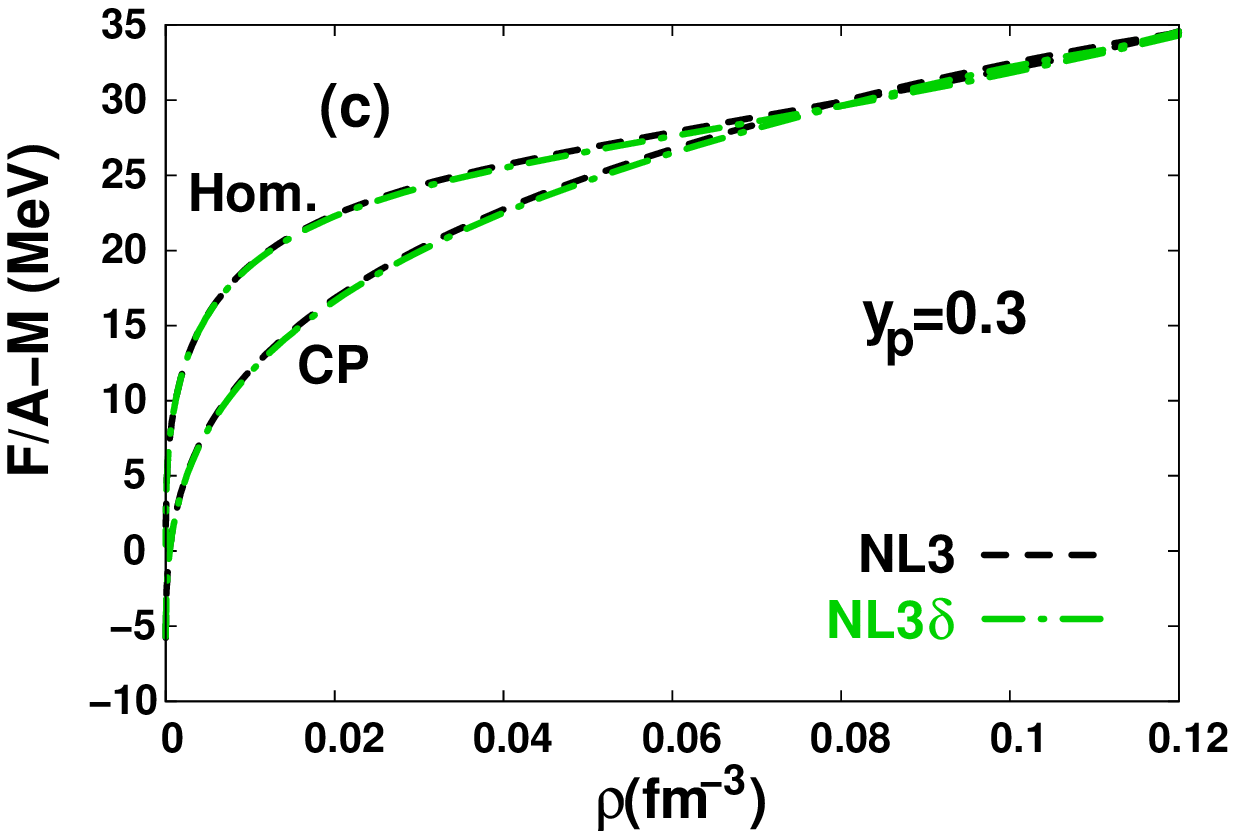} &
\includegraphics[width=4.5cm]{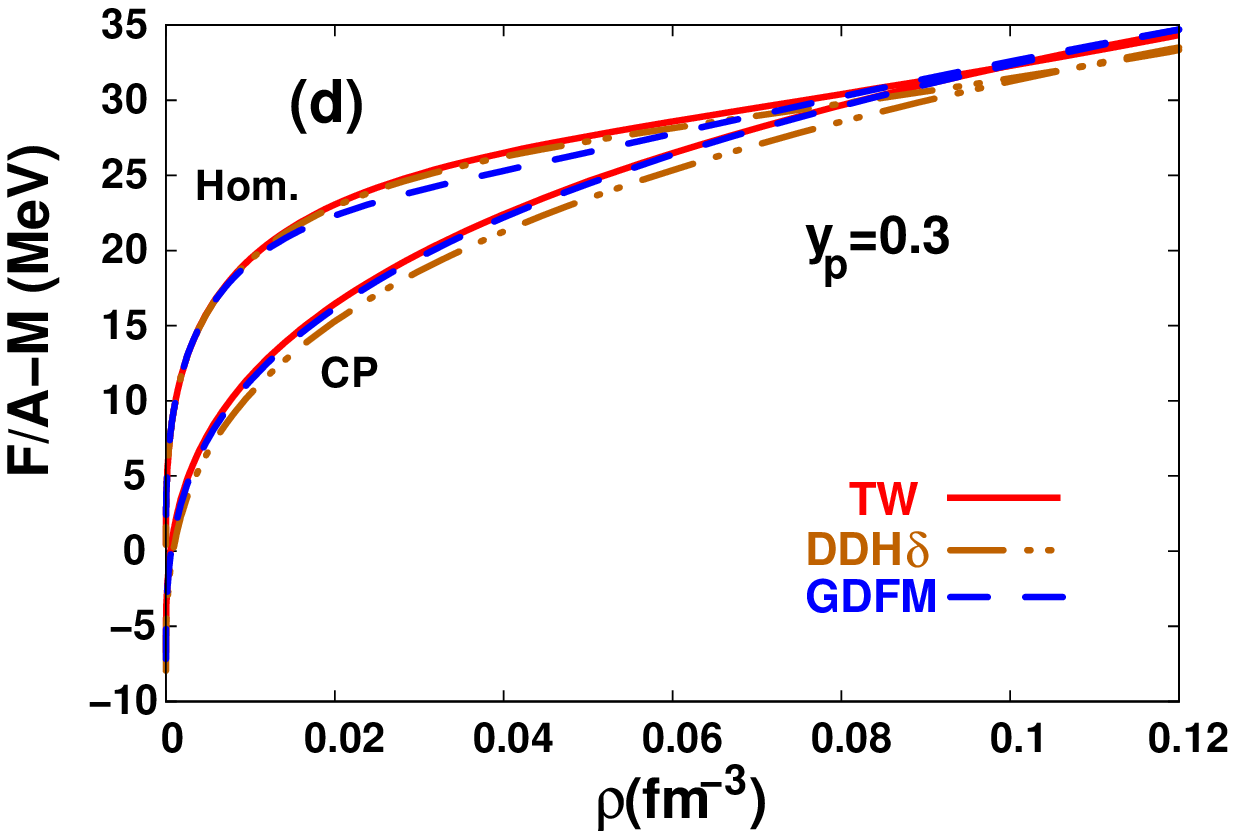}  \\
\end{tabular}
\end{center}
\caption{$npe$ matter (pasta + homogeneous phase) energy per particle at $T=0$ MeV for 
RMF models with constant couplings [a), c)] and density dependent coupling 
models [b), d)]: $y_p=0.5$ (top), $y_p=0.3$ (bottom).}
\label{fig2}
\end{figure}

\begin{figure*}[htb]
\begin{center}
\begin{tabular}{cc}
\includegraphics[width=7.cm]{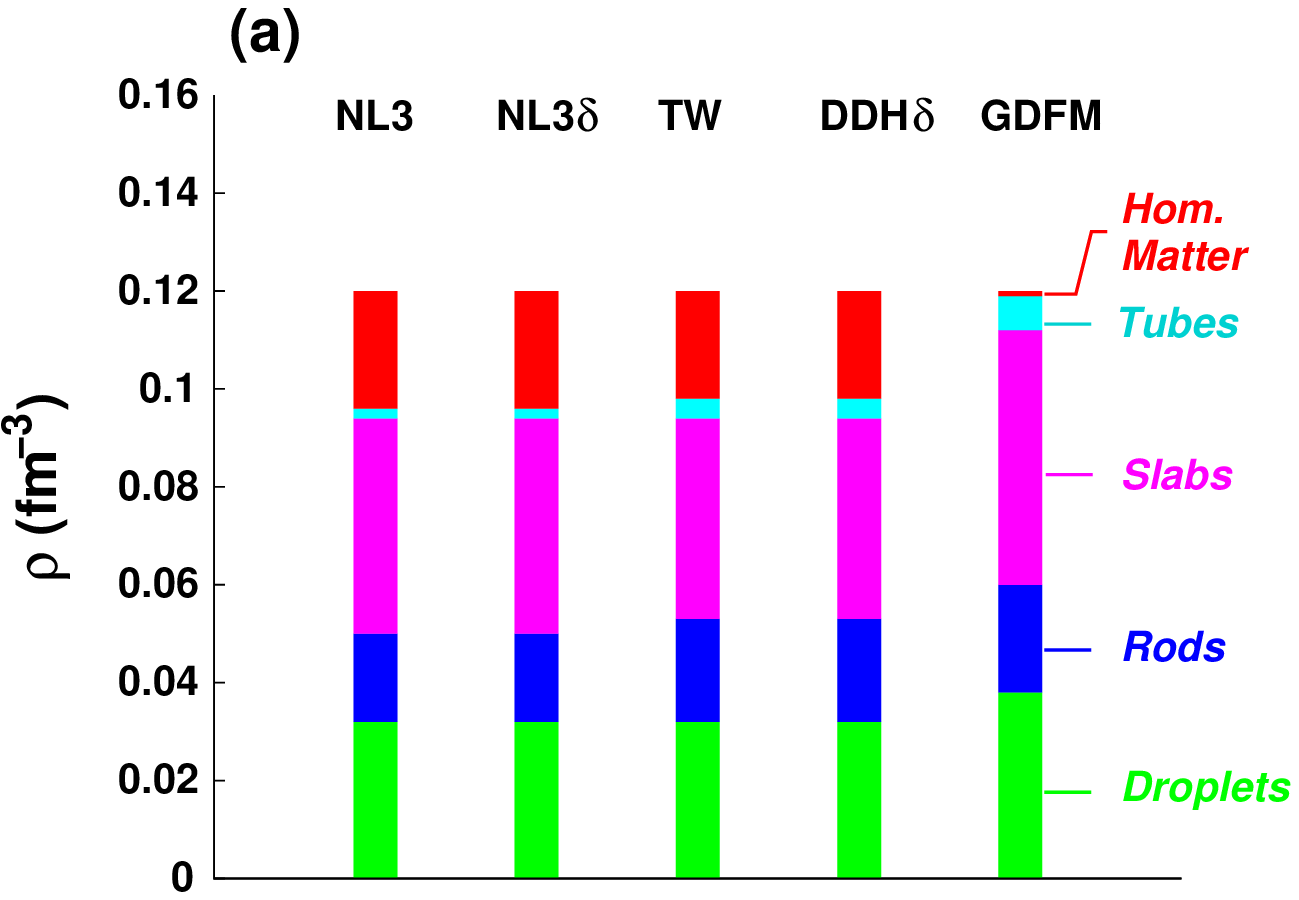} &
\includegraphics[width=7.cm]{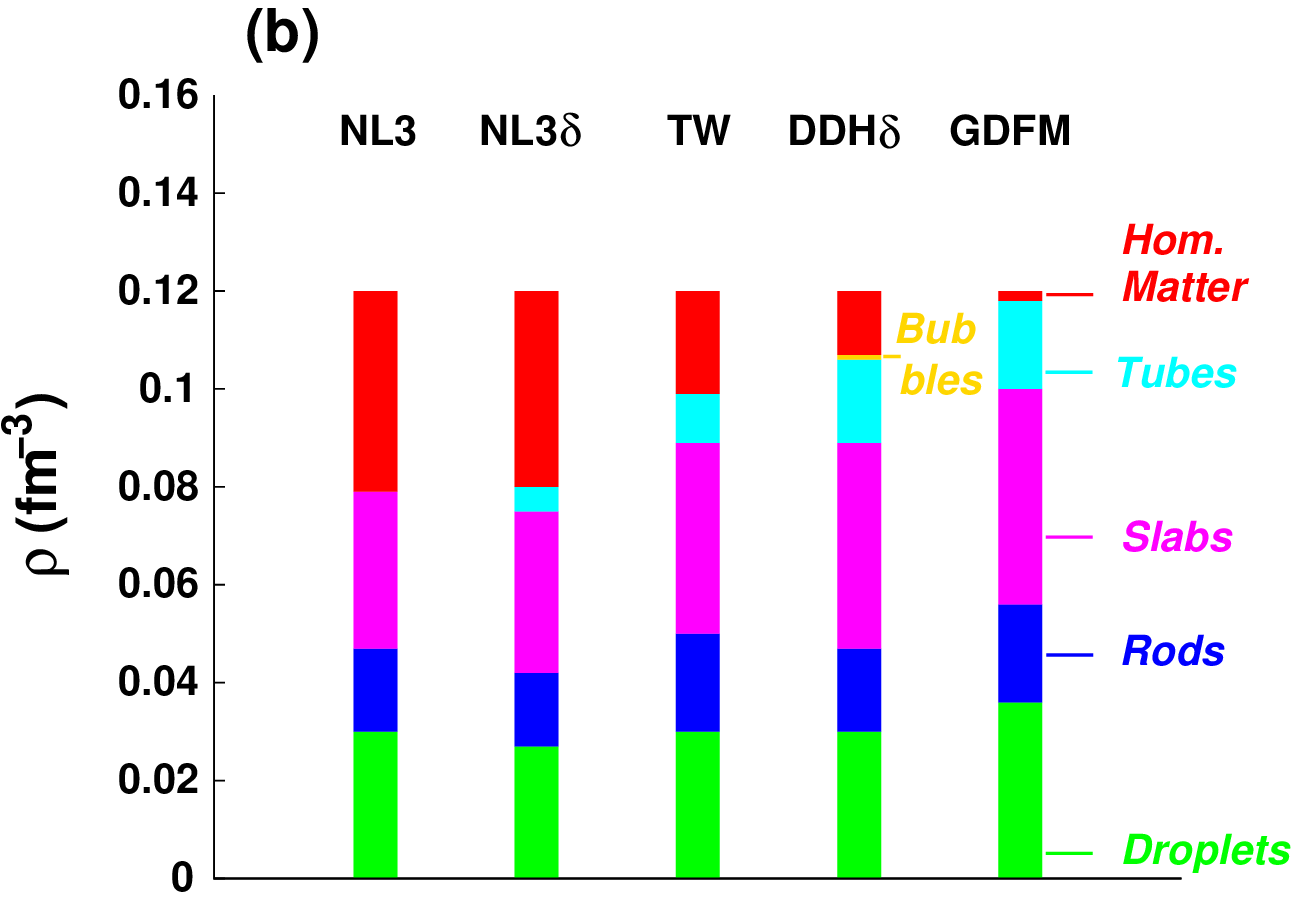} \\
\end{tabular}
\end{center}
\caption{Comparison of the phase diagrams at $T=0$ MeV for a) $y_p=0.5$ 
and b) $y_p=0.3$ obtained with the coexisting phases method (CP) 
for several models. From bottom to top the colors
represent droplets, rods, slabs, tubes and homogeneous matter.} 
\label{fig3}
\end{figure*}

In Fig. \ref{fig2} results for a homogeneous description of matter are compared with the output 
of the CP calculation for two proton fractions. We see that all 
the models predict the existence of a non-homogeneous pasta phase: in models 
with constant couplings, NL3 and NL3$\delta$, this phase clearly decreases 
if  $y_p$  decreases (higher asymmetries); a different behavior 
occurs for density dependent coupling models where the non-homogeneous phase 
extends to higher densities and may even increase
when the proton fraction reduces. From Table \ref{tzero} one can see that  
the TF method also predicts for density dependent coupling models either a small increase
(DDH$\delta$) or just a small decrease (TW) of the extension of the
pasta phase, if we decrease the proton fraction from 0.5 to 0.3.
The GDFM model presents a large pasta phase at both
asymmetries but it is with the DDH$\delta$ model that most variety of pasta
structures can be observed (see Fig. \ref{fig3}). 

\begin{figure}[bht]
\begin{center}
\includegraphics[width=7.cm]{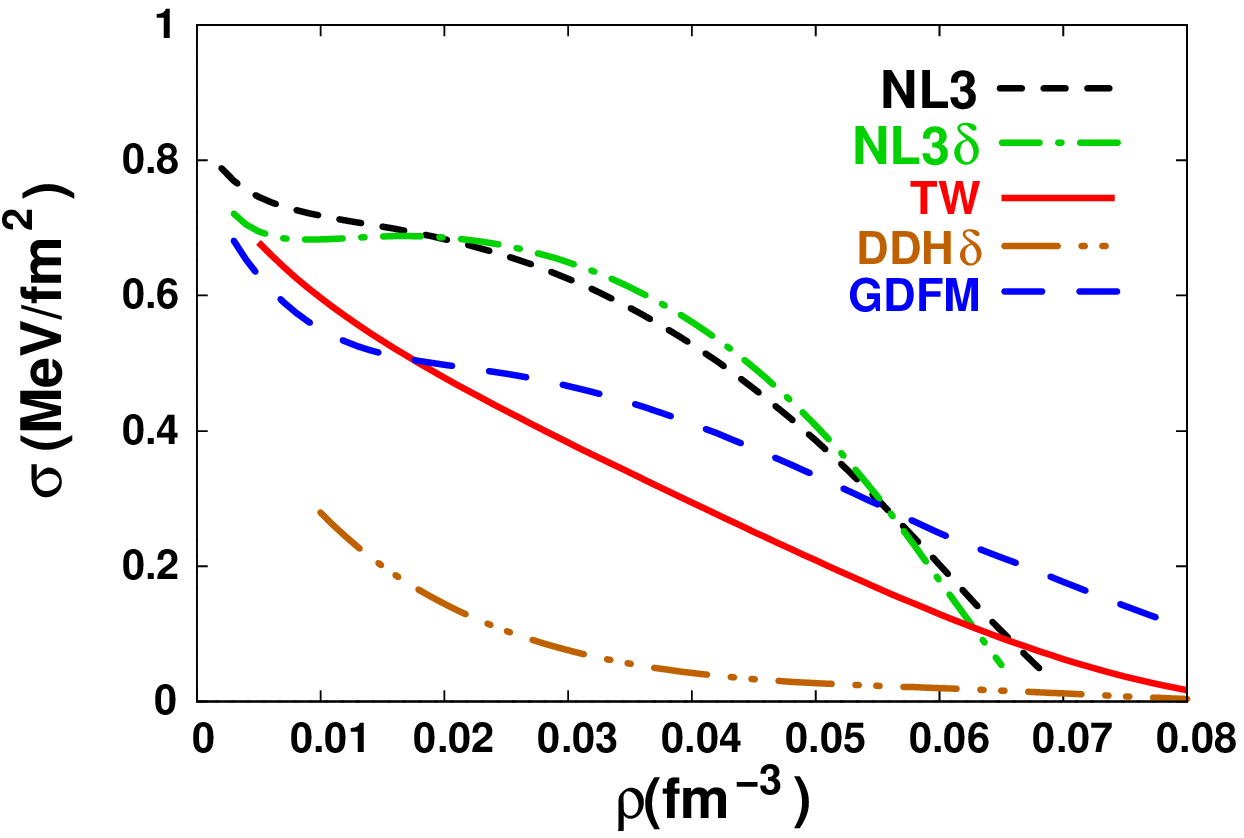} \\
\end{center}
\caption{Surface tension coefficient $\sigma$ at  $T=0$ MeV calculated according to eq. (\ref{sigma}). }
\label{fig4}
\end{figure}

\begin{figure}[bht]
\begin{center}
\includegraphics[width=7.cm]{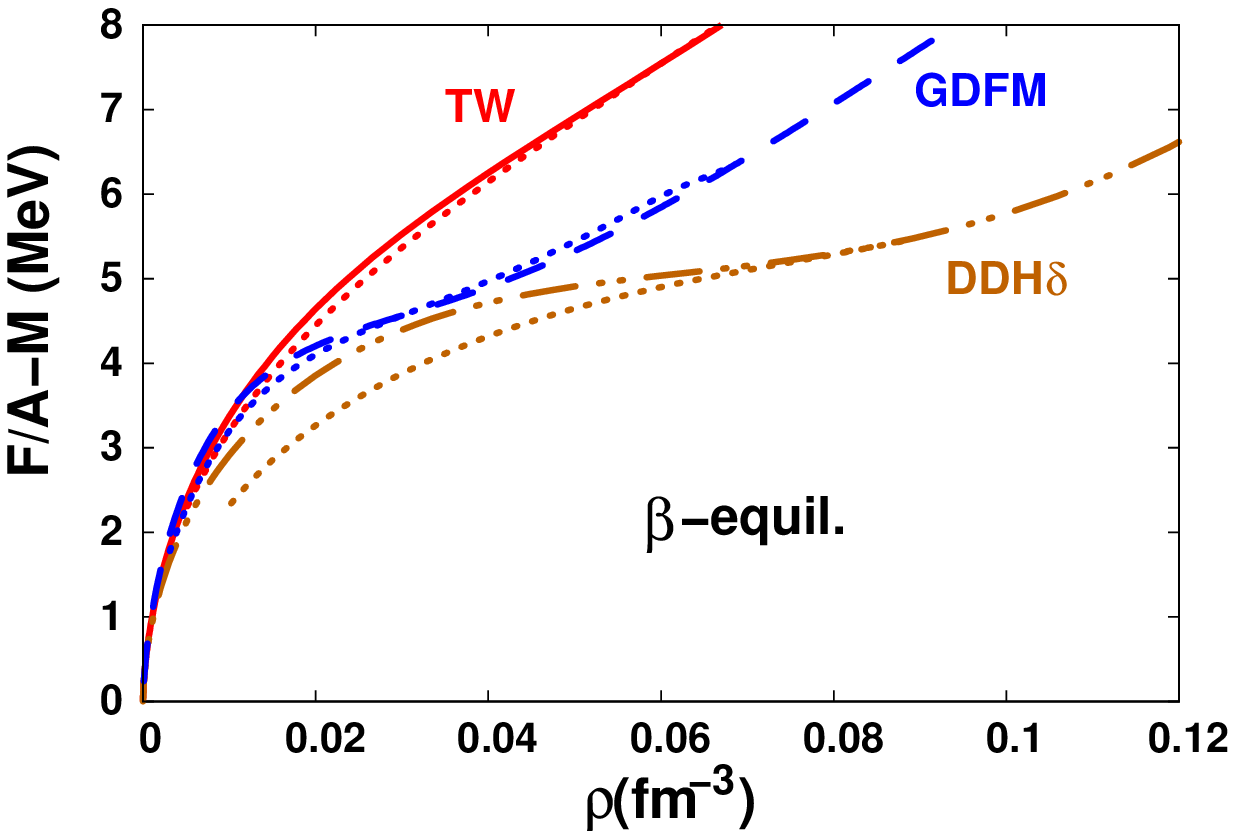} \\
\end{center}
\caption{$npe$ matter energy per particle at $T=0$ MeV for DDH models
at $\beta$ equilibrium. The full, dashed and dot-dot dashed lines stand for 
homogeneous matter and the corresponding dotted lines for pasta phases.}
\label{fig5}
\end{figure}

We also point out that the models NL3 and NL3$\delta$ as well as TW and 
DDH$\delta$ show the same transition densities for $y_p=0.5$ because the 
effect of the $\delta$ meson is only seen for asymmetric matter for the CP 
and spinodal calculations. Both couples of models have the same coupling constants for the 
$\sigma$ and the $\omega$ mesons and only differ in the isovector channel, namely the 
$\rho$ and $\delta$ meson couplings.
 However, in the TF calculation the distribution of protons and neutrons is free 
to adjust itself to the lowest energy configuration. As a consequence the proton and neutron 
density distributions do not coincide within the Wigner Seitz cell and the $\rho$ and $\delta$ 
fields are not zero. However, the differences are  not large enough to change 
the transition density and we still get the same transition density within TF, for $y_p=0.5$ and 
the couples  (NL3, NL3$\delta$) and (TW, DDH$\delta$).

For matter in beta-equilibrium  at $T=0$ MeV the energy per particle for the pasta is
always slightly larger than the corresponding homogeneous matter in the models
with constant couplings within the CP method; so, in these cases, the pasta 
is never preferred to represent the
ground state of the system. The absence of a pasta phase in the NL3 and NL3$\delta$ 
parametrisations is related to the very high values of the surface tension coefficient $\sigma$ for 
these models as can be seen in Fig. \ref{fig4}.

\begin{figure}[htb]
\begin{center}
\includegraphics[width=7.cm]{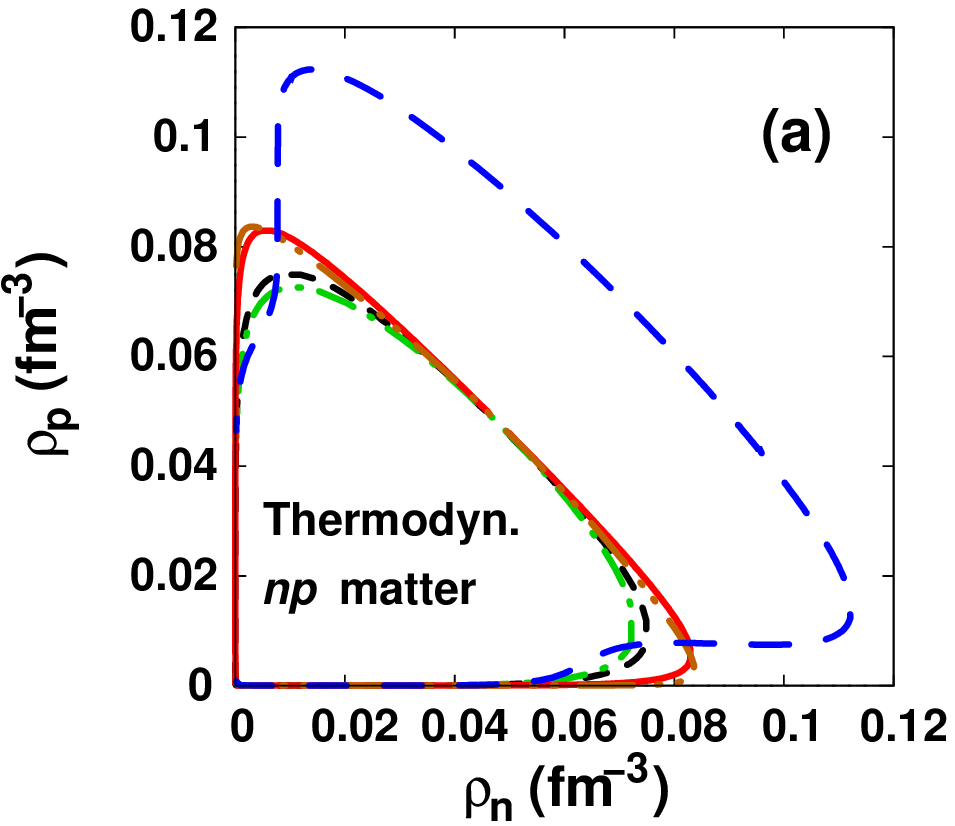} \\
\includegraphics[width=7.cm]{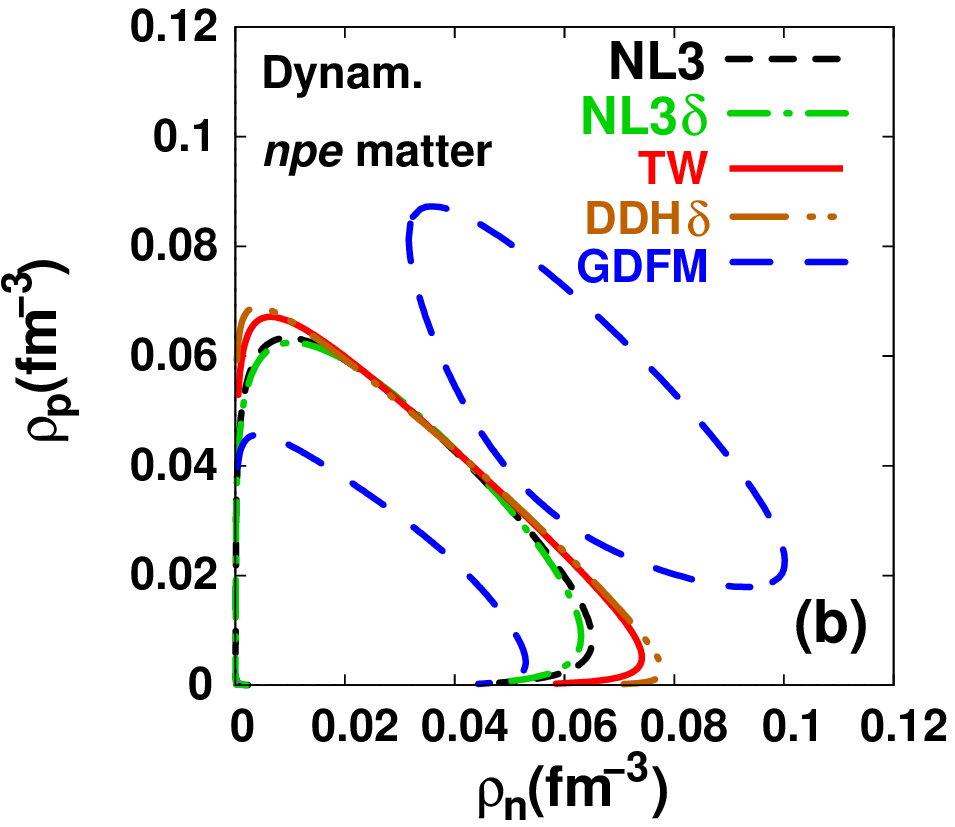} \\
\end{center}
\caption{a) Thermodynamical spinodal for $np$ matter  and b) dynamical spinodal 
for $npe$ matter, for a transfer  momentum $k=80$ MeV, for all the models under study. 
}
\label{fig6}
\end{figure}

For the DDH models we show in Fig. \ref{fig5} that these models 
predict non-homogeneous
phases occurring for a relatively small range of densities (below $\rho \sim
0.027$ fm$^{-3}$) for the GDFM model but extending up to  $\rho \sim
0.09$ fm$^{-3}$ in DDH$\delta$. TW shows an intermediate behavior. Due to the low symmetry
energy of the DDH$\delta$ model, which is only 25 MeV at saturation density,
it was shown in \cite{aziz08} that $\beta$-equilibrium matter would even
present at low densities a range of densities with a negative
compressibility. This behavior is favoring the appearance of a larger
variety of cluster forms.

Surface tension depends on the proton fraction of the high density region
inside the Wigner-Seitz cell. Models with a larger symmetry energy give a
dense region with a larger proton fraction and therefore a larger surface
energy (see eq. (\ref{sigma})). We also confirm 
that DDH$\delta$ shows the smallest values for this quantity.
Within the TF method, on the other hand, the pasta phase is also found
with the models bearing constant couplings. This is due to the fact that 
in the TF approach the surface energy is calculated self-consistently and is
not introduced by hand. We know, however, that the TF approach predicts a too
steep surface and, therefore, we may expect that a quantal approach would
predict a larger pasta phase \cite{peles}. 

The spinodal surface gives information about the minimal dimension  of the 
pasta phase. A spinodal decomposition would be expected in the case of a fast 
transition, however in stellar matter we may expect that there is always 
plenty of time to get equilibration.  In \cite{bao} it was shown that the 
thermodynamical spinodal results for $pn$ matter did not differ very
  much from the dynamic spinodal ones for $npe$ matter. This seems to 
indicate that the
  Coulomb interaction and surface tension do not influence a lot the pasta
  phase extension. The  thermodynamical spinodal for
  $npe$ matter  either does not exist for density dependent hadronic 
models or is very small
  for NLW models due to the large incompressibility of electrons. However, 
although
  thermodynamically stable, $npe$ matter clusterizes as soon as it suffers a
  density fluctuation due to any kind of perturbation.  Therefore we 
discuss the extension of the non-homogeneous phase by analyzing the 
dynamical spinodal for $npe$ matter, within the several models considered.

 In Fig. \ref{fig6} we display both the thermodynamical
spinodals for $np$ matter  and the dynamical
spinodals for $npe$ matter for a momentum transfer $k=80$ MeV, which essentially
defines the envelope of the spinodal surfaces for all $k$ values 
\cite{umodes06}. The dynamical spinodals are smaller 
than the thermodynamical ones as expected. 
The crossing density of the EoS with a fixed proton fraction equal to 0.5 and 0.3 or 
for $\beta$-equilibrium matter is given in Table \ref{tzero}.

A larger extension of the pasta phase within the
GDFM model is expected from the thermodynamical spinodal which we show in
Fig. \ref{fig6}.
We notice that GDFM has a very 
peculiar behavior with a much larger thermodynamical spinodal. There is, 
however, an intermediate density region where matter is not
so unstable and the presence of electrons is enough to raise the instability
giving origin to two disconnected unstable regions. Comparing all the
spinodals we expect smaller non-homogeneous regions for NL3$\delta$ and a larger
one for GDFM if $y_p$ is not too small. For very asymmetric matter 
like matter in $\beta$-equilibrium, the TW and DDH$\delta$ models bear the largest 
pasta phases and NL3, NL3$\delta$ and GDFM predict  
similar results. We also verify that the dynamical spinodal predicts a slight 
increase of the unstable region when $y_p$ decreases from 0.5 to 0.3. 
This behavior is directly related with the concavity of the spinodal at $y_p=0.5$. 
In  \cite{camille08} it was seen that the concavity of the thermodynamical spinodal 
for the TW parametrization at $y_p=0.5$ is smaller than the one obtained with 
the NLWM.  The presence of electrons and the Coulomb field in the calculation of the 
dynamical spinodal gives rise to a spinodal that is not symmetric with respect to the $y_p=0.5$ axis. 
The spinodal may extend to larger densities for smaller proton fractions and the same isospin asymmetry.  
As discussed in \cite{chomaz07} we also expect a larger extension of the non-homogeneous 
phase if the electron contribution is described correctly, and the Coulomb field included self consistently,
what stabilizes $npe$ matter and extends the non-homogeneous phase.

It is interesting to compare  the density transitions obtained within 
the spinodal
  approaches with the corresponding values determined from the minimization of
  the free energy both within the CP and TF approaches. As discussed before,
  the dynamical spinodals are expected to indicate a lower limit.
 Within the present models, the transition densities obtained from the 
thermodynamical spinodal are $\sim$ 10-15\% larger than the values obtained 
from the dynamical
  spinodal, similar to the conclusion drawn in \cite{bao}.
For the proton fractions 0.5 and 0.3 these values are always larger than the
ones obtained within an equilibrium calculation, either CP or TF, except for
the GDFM model for which the CP results are smaller than the spinodal
ones. For the  $\beta$-equilibrium calculation the CP method predicts  
no pasta phase for the NLWM parametrisations (NL3 and
NL3$\delta$). This is due to the non self-consistent description of the 
surface in the CP approach.
 The TF approach, which treats self-consistently the surface, 
 predicts, for all models, a  transition density larger, but very
 similar, to the one predicted by the
 dynamical spinodal calculation. This result is very interesting because it
 implies that to calculate the transition density at the inner edge of the
 compact star crust it is enough to use a dynamical spinodal calculation.

In Table \ref{tdens} we
compare the transition densities between the different pasta
geometries obtained in the present calculation with the 
results from \cite{gogelein}, where both a TF and a microscopic calculation 
were done. 
In the TF calculation the
surface description was not fully self-consistent because it involved 
the inclusion of a surface energy parameter that
was adjusted to reproduce the experimental binding energy of the
nucleus $^{208}$Pb. We conclude that with a self-consistent TF
calculation the transition densities between the different geometries
and from non-homogeneous to homogeneous matter are quite smaller than
the results in \cite{gogelein} obtained within a microscopic
description of the pasta structures including pairing effects. This
comparison should also be done for different proton fractions and not only
for $\beta$-equilibrium matter.

\begin{table}[b]
\caption{Transition densities in fm$^{-3}$ between the different 
geometries at $T=0$ MeV and
  for the GDFM model. Comparison with the results taken from \cite{gogelein}.}
\begin{tabular}{ccccc}
\hline
 & CP& TF & TF \cite{gogelein}&H \cite{gogelein}\\
\hline
droplet-rod& &0.047 & 0.048& 0.052\\
rod-slab & & 0.048& - & -\\
slab-hom.& 0.027 & 0.052 & 0.061& 0.064\\
\hline
\end{tabular}
\label{tdens}
\end{table}

\begin{figure}[t]
\begin{center}
\includegraphics[width=7.cm]{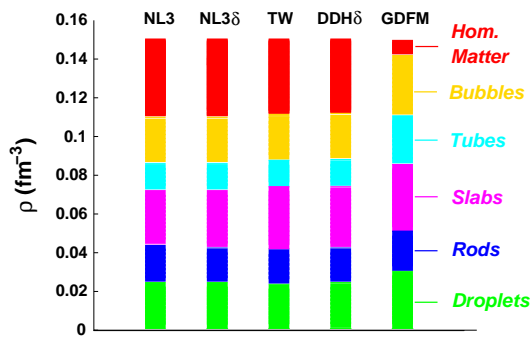}
\end{center}
\caption{Phase diagrams at $T=0$ MeV and $y_p=0.5$ obtained with the 
TF method for several models. From bottom to top the colors
represent droplets, rods, slabs, tubes, bubbles and homogeneous matter.} 
\label{fig7}
\end{figure}

>From the phase diagrams shown in  Figs. \ref{fig3}a) and \ref{fig3}b) 
we observe that  most models predict the formation of inhomogeneities of
the type droplet, rod, slab and tube for the asymmetries considered. For
$y_p=0.3$ only NL3 evolves to homogeneous matter without a tube-like
structure and, on the other side, DDH$\delta$ predicts the appearance of
bubbles in a narrow interval of densities. These differences are due to
  the dependence of the surface energy on the proton asymmetry and on the
  slope of the symmetry energy. In Fig. \ref{fig7} the phase diagrams 
obtained with the TF method for the models under study are displayed with
$y_p=0.5$. In this case the bubble structure, not present in
Fig. \ref{fig3}a), appears. The transition densities are systematically
higher with TF than with CP, as seen in Table \ref{tzero}, and, therefore another phase 
structure is accommodated.

\begin{figure}[b]
\begin{center}
\includegraphics[width=7.cm]{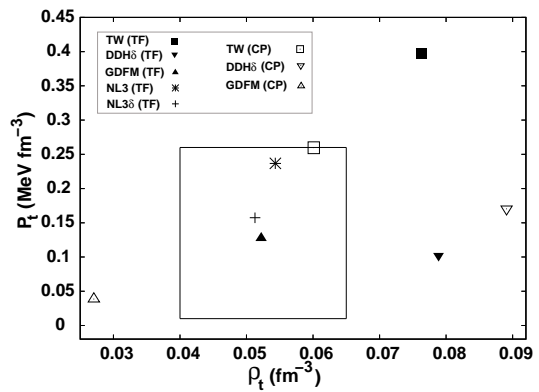}
\end{center}
\caption{Density and pressure of the inner edge that divides the liquid core 
from the solid crust of neutron stars according to \cite{bao}.}
\label{fig8}
\end{figure}

At this point we compare our results shown in Table \ref{tzero} for $npe$ 
matter in $\beta$-equilibrium with the
predictions given in \cite{bao}. According to \cite{bao} the transition 
density and pressure from the liquid core to the solid crust at the inner edge 
of neutron stars should lie within the ranges 
\beq
0.04 \le \rho \le 0.065 \mbox{ fm}^{-3}, \qquad 0.01 \le P \le 0.26
\mbox{ MeV/fm}^3.\label{rprange}
\eeq
The values for the pressure 
were constrained to the values for the slope of the symmetry energy at the saturation density given
by \mbox{ $L=86 \pm 25$ MeV.} The limits on the transition pressure defined in 
\cite{bao}
are, however, quite smaller than the ones given in \cite{link99}, namely
$ 0.25 \le P \le 0.65 \mbox{MeV/fm}^3$, obtained from realistic EoS.

 One can see in Table \ref{prop} that the NL3 and 
NL3$\delta$ models have $L$ values respectively slightly larger and larger  
than the upper limit of the proposed $L$ values. However, using the
 TF method, we get, for the transition densities and related pressures from 
the pasta phase to the  homogeneous phase, values within the proposed range. 
On the other hand, the density dependent models have 
$L$ values slightly smaller than the lower limit of the above $L$ range.

The results for the transition pressure and density in the TW model lie 
just at the border of the rectangle that sets the limits given in 
eqs. (\ref{rprange}) for the CP
 calculation but they become too high when the TF method is used.
For the DDH$\delta$ parameter set, the results for 
the pressure are inside the range shown in eqs.(\ref{rprange}) both in the CP 
and TF approaches. However, for this model the transition density is too high 
for both calculations.
For the GDFM parametrisation the density within the
CP method is not good since it comes up too low, but the TF results lie within 
the imposed constraints. For the NL3 and NL3$\delta$ parametrisations, the
pasta phase is only obtained within the TF results and they come out inside
the constrained range. These observations are
 summarized in   Fig. \ref{fig8}. 

Within
the dynamical spinodal calculation, the transition densities
for TW and DDH$\delta$ are high, but all the other models are within the  
density range given in eqs. (\ref{rprange}).

Also from Table \ref{tzero} one observes that the influence of the $\delta$ 
meson is only effective for large proton asymmetries. For proton
fractions 0.5 and 0.3, NL3 and NL3$\delta$
give similar results. 
The effect of the $\delta$ meson is only observed in the
$\beta$-equilibrium matter results: for the constant coupling models the
inclusion of the $\delta$-meson makes the pasta phase range a bit smaller, the
results of the TF calculation being in good agreement with the dynamical
spinodal ones.
Among the density dependent models and considering all type of calculations
presented, we see that the extension of the pasta phase for DDH$\delta$ is
larger than the corresponding one within TW which does not include the 
$\delta$ mesons. 
This is due to the low value of the symmetry energy within the DDH$\delta$ 
model.

\begin{figure}[thb]
\begin{center}
\begin{tabular}{cc}
\includegraphics[width=4.cm]{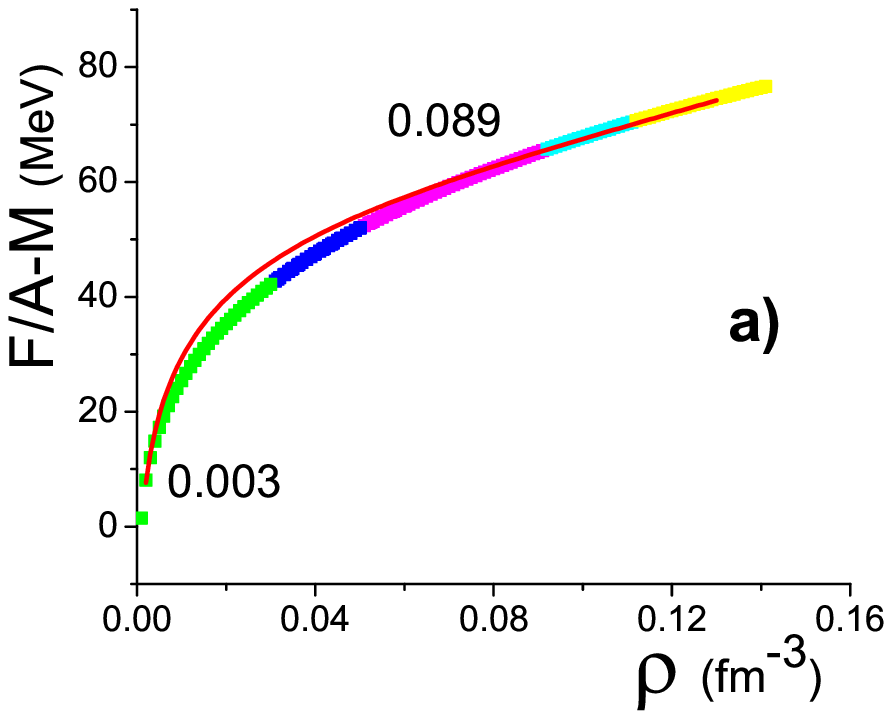} &
\includegraphics[width=4.0cm]{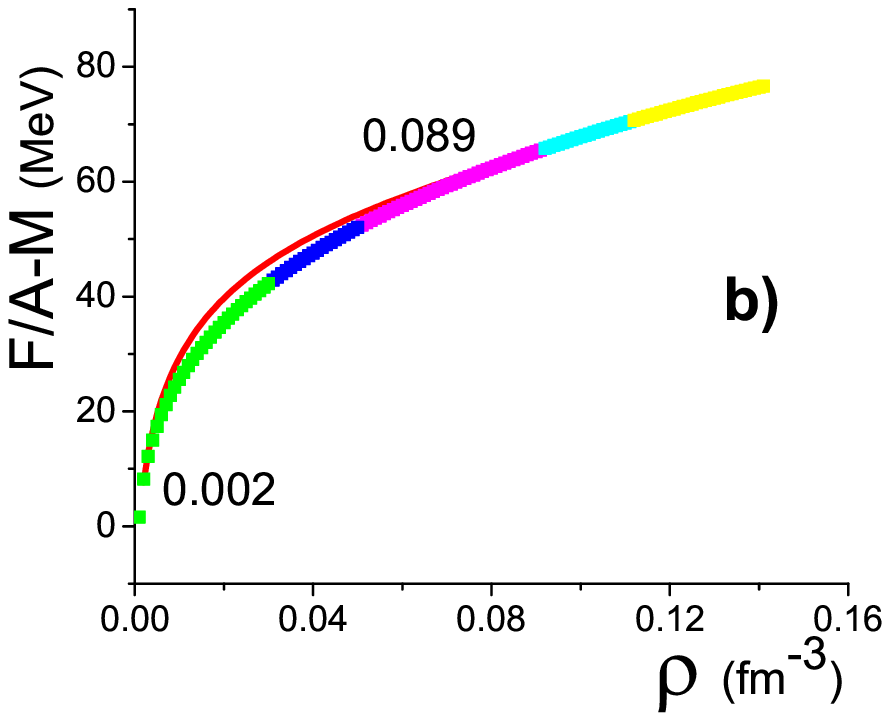} \\
 \includegraphics[width=4.0cm]{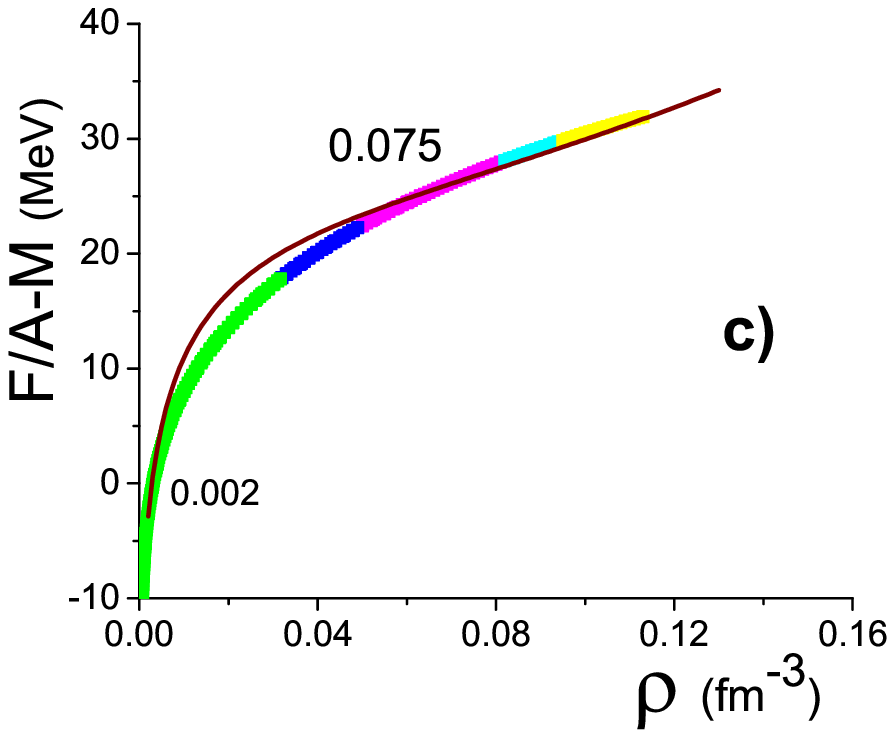} &
\includegraphics[width=4.0cm]{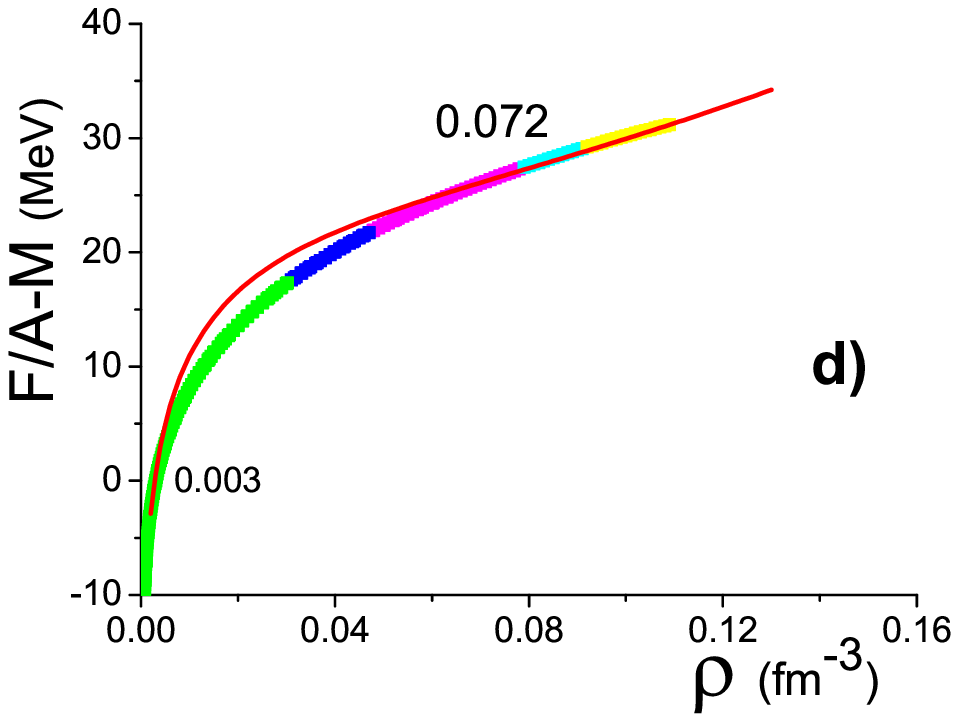} \\
\end{tabular}
\end{center}
\caption{$npe$ matter (pasta + homogeneous phases) free energy per particle at $T=5$ MeV 
for RMF models with constant couplings: NL3 - a) and c), NL3$\delta$ - b) and d). $y_p=0.5$ (top), $y_p=0.3$ (bottom).}
\label{fig9}
\end{figure}

We now comment on the results with finite temperature, all of them obtained
with the CP method. In Fig. \ref{fig9} we
plot the free energy per particle for the models with constant couplings. As 
expected from the $T=0$ MeV results, no pasta phase appears when 
$\beta$-equilibrium is enforced. For fixed proton fractions (0.5 and 0.3), 
the pasta phase shrinks with temperature.  At very low densities the homogeneous
phase has a lower free energy than the pasta phase. 

\begin{figure*}[htb]
\begin{center}
\begin{tabular}{ccc}
\includegraphics[width=5.5cm]{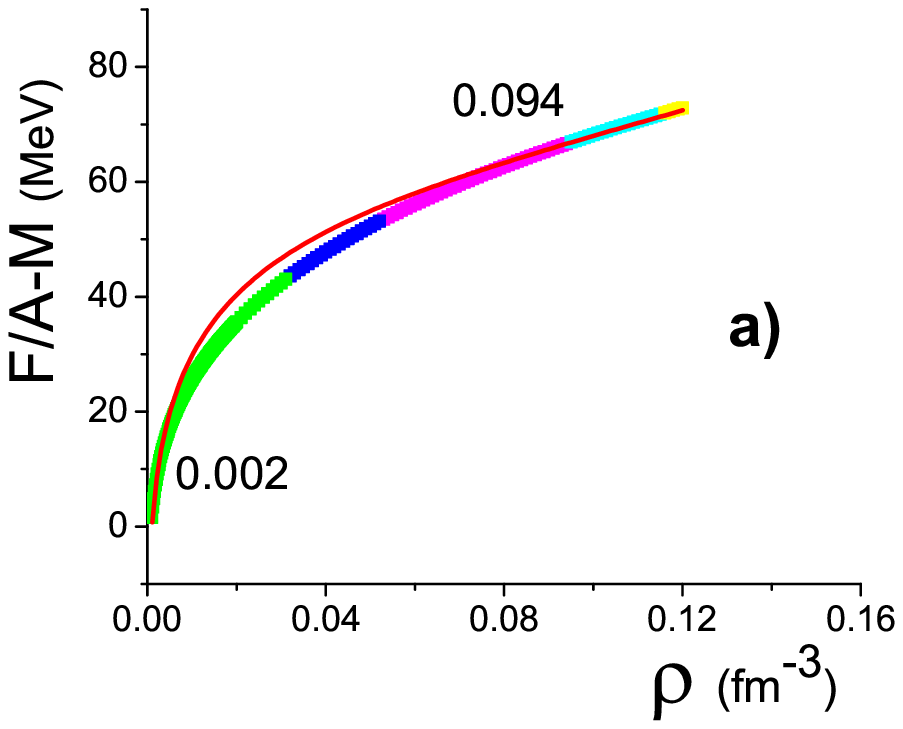} &
\includegraphics[width=5.5cm]{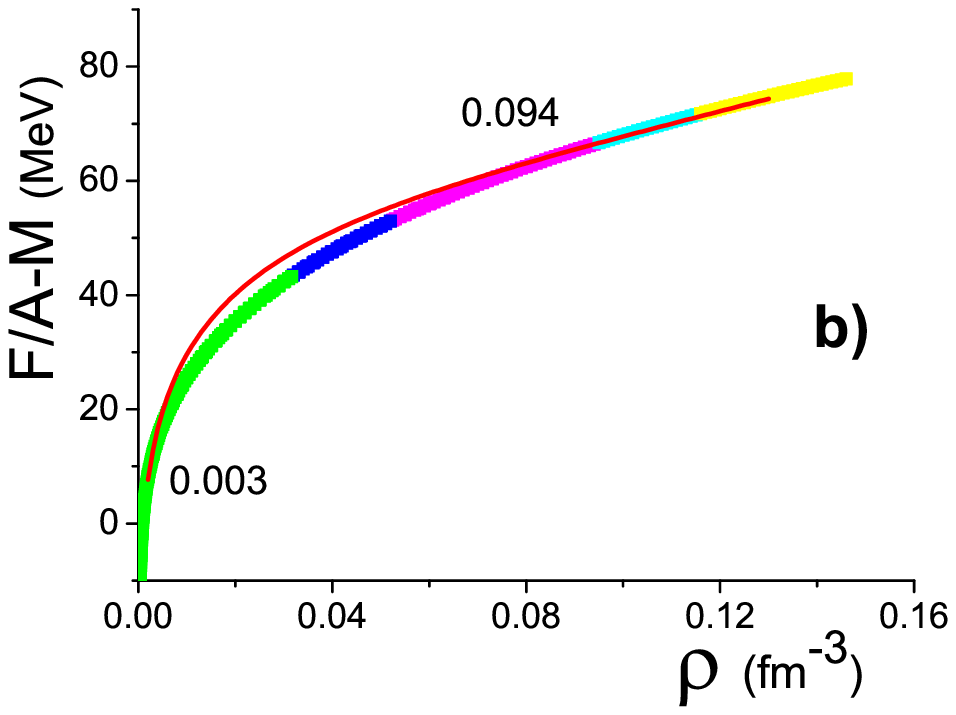} &
\includegraphics[width=5.5cm]{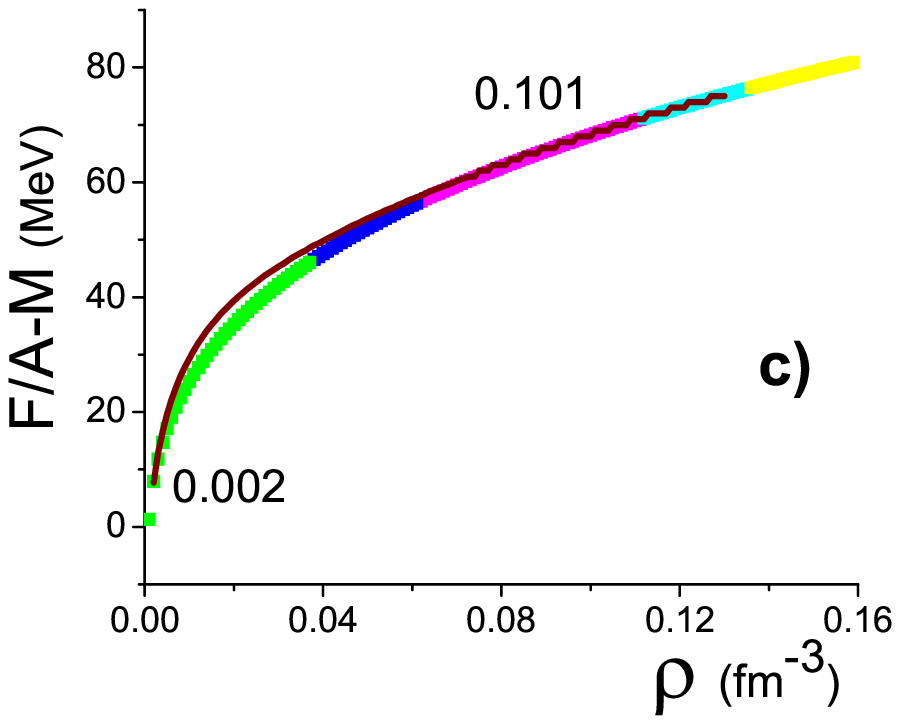} \\
\includegraphics[width=5.5cm]{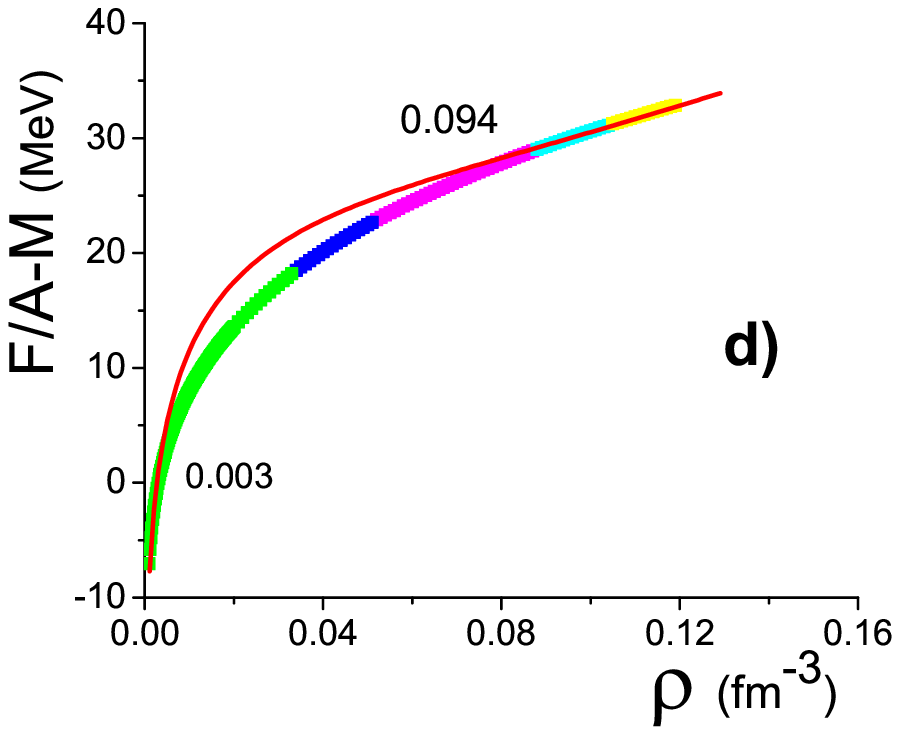} &
\includegraphics[width=5.5cm]{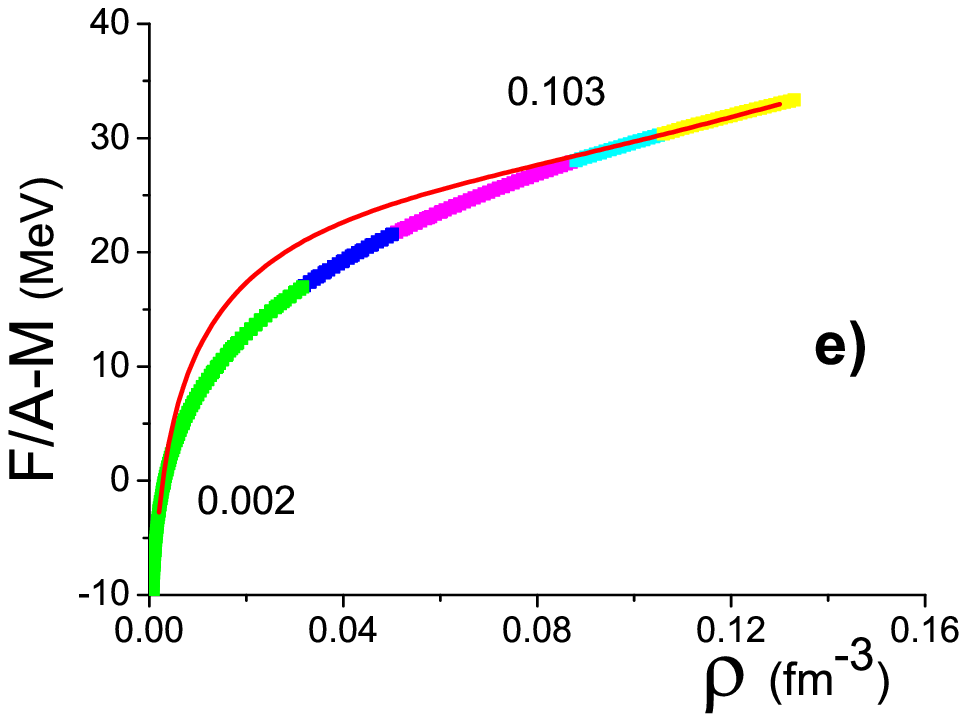}&
\includegraphics[width=5.5cm]{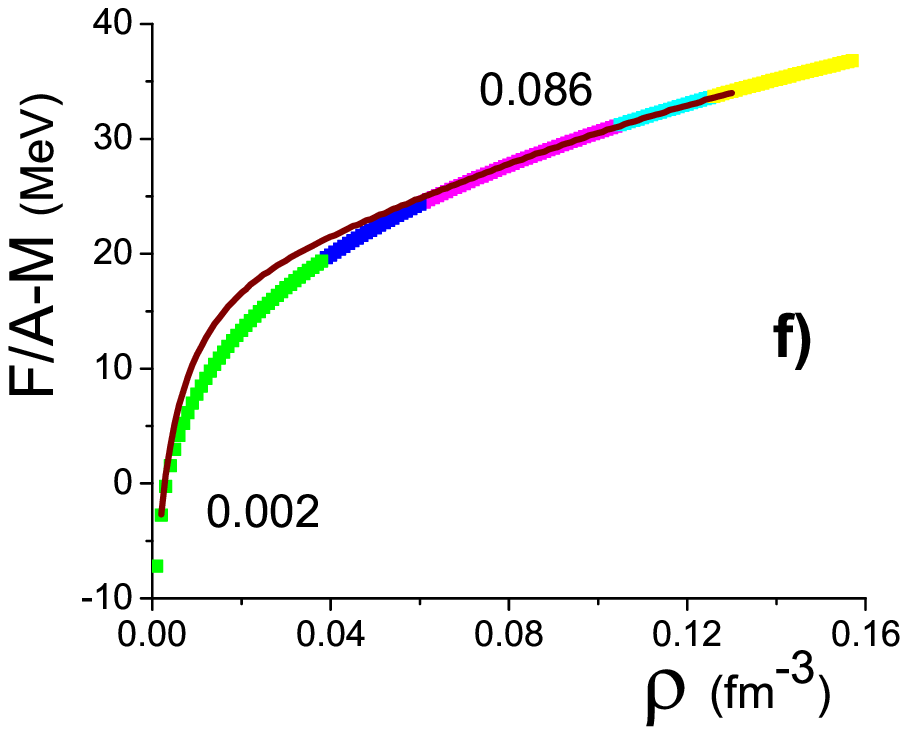} \\
\includegraphics[width=5.5cm]{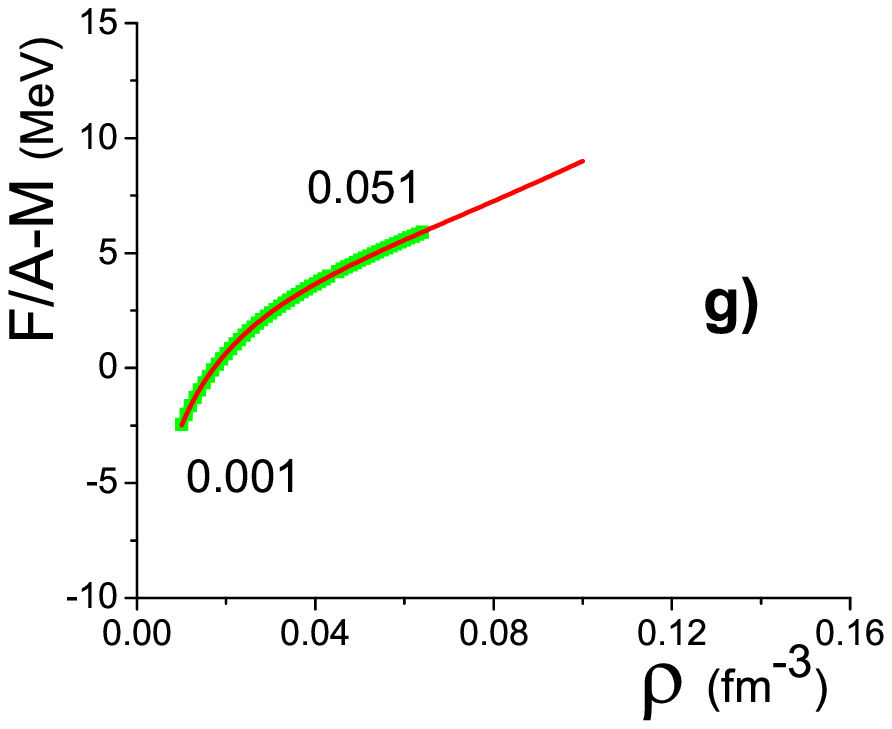} &
\includegraphics[width=5.5cm]{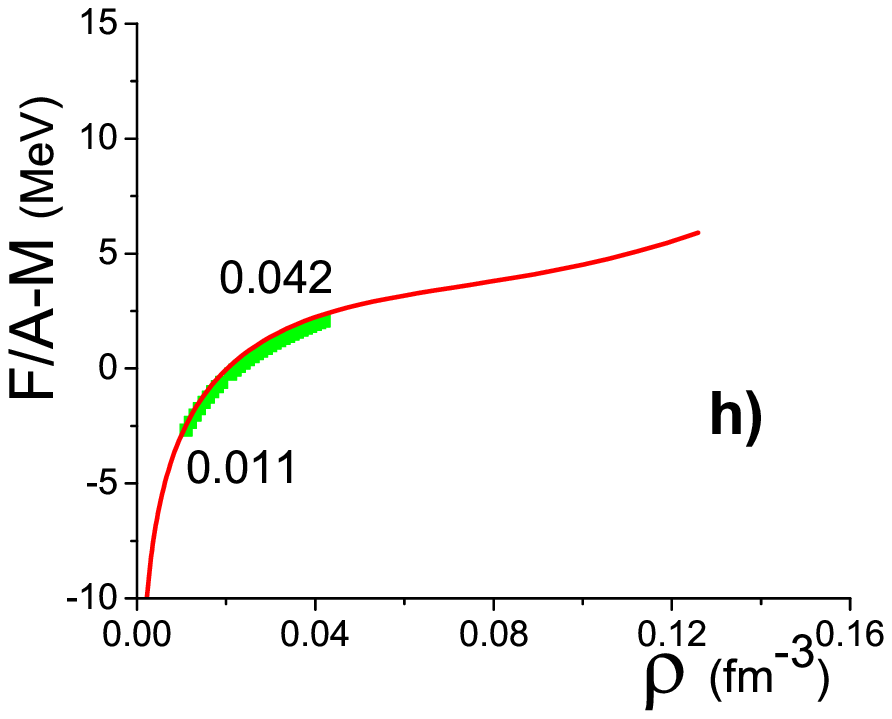}& \\
\end{tabular}
\end{center}
\caption{$npe$ matter (pasta + homogeneous phase) free energy per particle at $T=5$ MeV 
for RMF models with density dependent couplings: TW - a), d) and g), DDH$\delta$ -  b), e) and h), GDFM - c) and f).
$y_p=0.5$ (top), 
$y_p=0.3$ (middle) and $\beta$-equilibrium (bottom).}
\label{fig10}
\end{figure*}

\begin{figure*}[htb]
\begin{center}
\begin{tabular}{cc}
\includegraphics[width=7.cm]{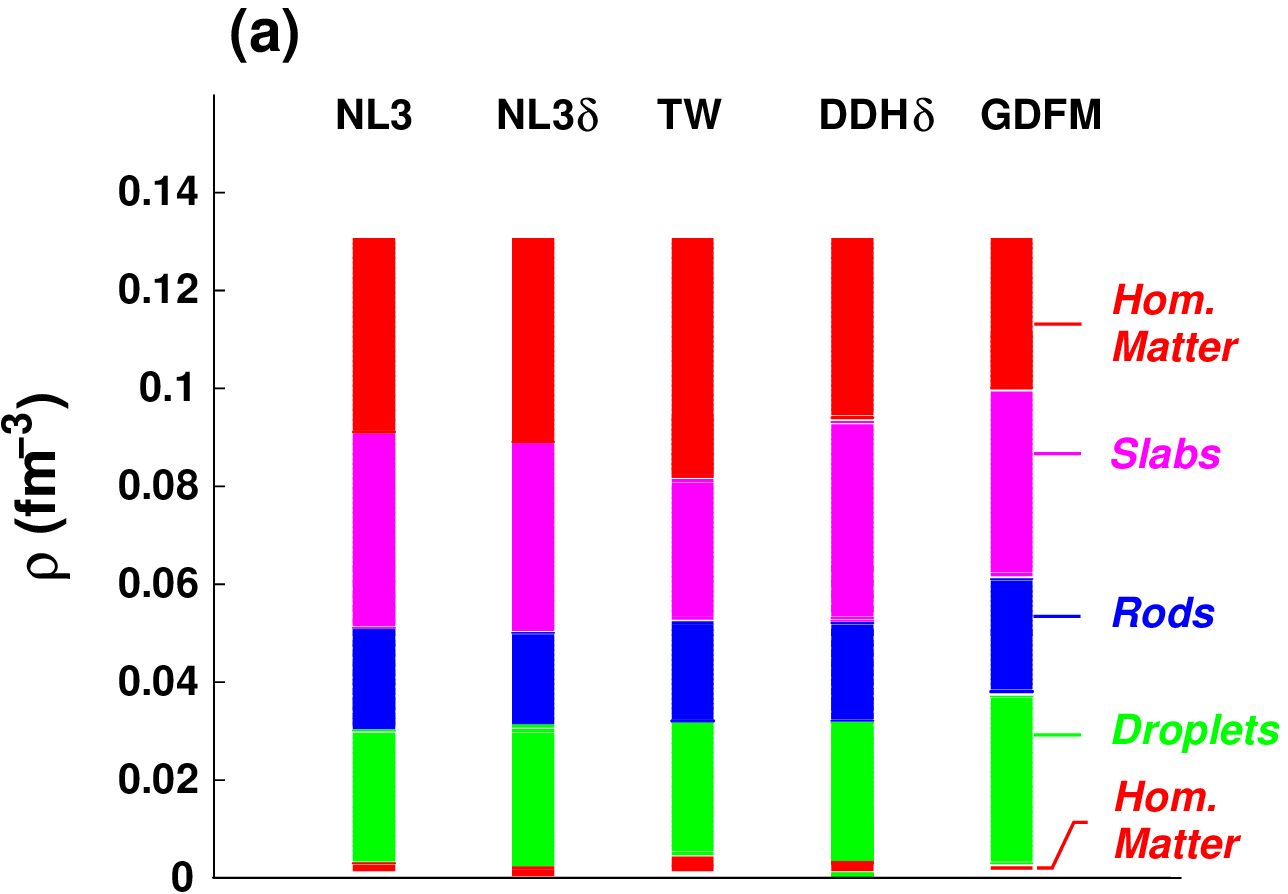} &
\includegraphics[width=7.cm]{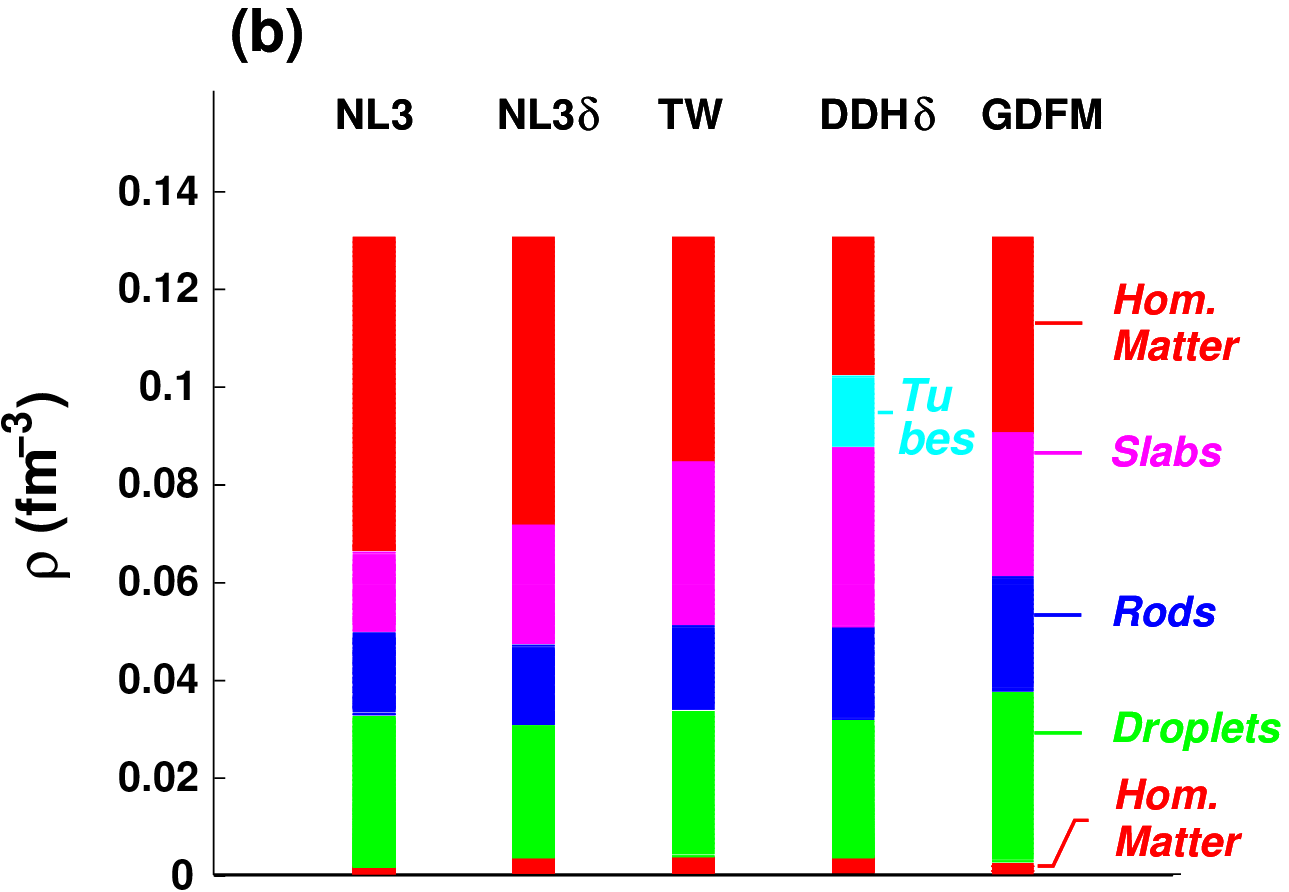} \\
\end{tabular}
\end{center}
\caption{Comparison of the phase diagrams at $T=5$ MeV for a) $y_p=0.5$ 
and b) $y_p=0.3$ obtained with the coexisting phases method (CP) 
for several models. From bottom to top the colors
represent homogeneous (for some cases only), droplets, rods, slabs, tubes and 
homogeneous matter.} 
\label{fig11}
\end{figure*}

This behavior had already been noticed in \cite{pasta1} and is reproduced with
DDH parametrisations, as seen in Fig. \ref{fig10} for $y_p=0.5, \, 0.3$ and for
matter in $\beta$-equilibrium. From Fig. 
\ref{fig10} and \ref{fig11} it is seen that the size of the pasta phase depends 
on the asymmetry of the $npe$ matter and on the chosen parametrisation. Just 
two models provide pasta phase within CP for matter in $\beta$-equilibrium at 
$T=5$ MeV: TW and DDH$\delta$, the second being larger than the first.

\begin{table}[b]
\caption{Transition densities in fm$^{-3}$ and corresponding pressures (CP calculation) for 
the non-homogeneous to homogeneous phase at the inner edge of the crust ($T=5$ MeV).}
\label{t5}
\begin{center}
\begin{tabular}{cccccccc}
\hline
model & EoS &  thermodynamical & pasta (CP)  & P (CP)\\
      &     &   spinodal  &     versus & MeV/fm$^3$ \\
      &     &  versus EoS & EoS & uniform matter & \\
\hline
NL3  & $y_p=0.5$       &  0.094   & 0.089  & 2.42    \\
NL3  & $y_p=0.3$       &  0.090  & 0.075  &  1.02\\
NL3  & $\beta$-equil. &  -  & -  & & \\
\hline
NL3$\delta$  & $y_p=0.5$       & 0.094 &   0.089  &  2.41 \\
NL3$\delta$  & $y_p=0.3$       &  0.090 &  0.072   &  0.96\\
NL3$\delta$  & $\beta$-equil. &  -  &  -  & \\
\hline
TW  & $y_p=0.5$       & 0.095  &  0.094 & 2.59 \\
TW  & $y_p=0.3$       & 0.094  & 0.094  & 0.94 \\
TW  & $\beta$-equil. &  0.051  & 0.035  & 0.15 \\
\hline
DDH$\delta$  & $y_p=0.5$       & 0.095  & 0.094 & 2.59\\
DDH$\delta$  & $y_p=0.3$       & 0.093  & 0.103 & 1.51\\
DDH$\delta$  & $\beta$-equil. &  0.073  &0.042  & 0.10\\
\hline
GDFM  & $y_p=0.5$       & 0.140  &  0.101  & 2.84\\
GDFM  & $y_p=0.3$       & 0.137  & 0.086  & 1.20\\
GDFM  & $\beta$-equil.  &  0.029  & -  & - \\
\hline
\end{tabular}
\end{center}
\end{table}

\begin{table}[bth]
\caption{Highest temperatures for which pasta phase was found in MeV.}
\label{tabtemp}
\begin{center}
\begin{tabular}{ccccccccc}
\hline
&  NL3 & NL3$\delta$ & TW   & DDH$\delta$ &GDFM \\
&  \cite{nl3} &  & \cite{tw} &  \cite{gaitanos} & \cite{gogelein}\\
\hline
\hline
$y_p=0.5$ &12.0 & 12.0 & 14.0 & 14.0 & 10.0\\
$y_p=0.3$ &10.0 & 7.0 & 13.0 & 12.0 & 8.0\\
$\beta$-equil.& - & - & 5.0& 8.0& 4.0 \\   
\hline
\end{tabular}
\end{center}
\end{table}

In Table \ref{t5} the transition densities obtained within the CP
approach and using the thermodynamical spinodal are given. No data 
for a dynamical spinodal calculation at  finite temperature is available, 
except for NL3, see \cite{pasta1}, where the transition densities 
0.080, 0.077 fm$^{-3}$ were given respectively for $y_p=0.5$ and 0.3 at $T=5$ 
MeV. In this case no crossing occurs for $\beta$-equilibrium matter.   
These values are similar to the ones obtained within the CP calculation.
As discussed before we expect that the
dynamical spinodal defines a lower limit for the transition
density. From the discussion of the results obtained for the pasta phase 
at $T=0$ MeV, we also expect that the CP calculation only gives a lower limit 
for the transition density  because of the description of the
  surface which takes too large values.
 The thermodynamical spinodal only suggests an order of
magnitude, which, according to \cite{pasta1} and for NL3,  
was always a bit larger than the values coming from the CP approach and 
closer to the TF results for the NLW models.

For the DDH models, it is seen that within the 
CP approach, and similarly to the result already discussed at $T=0$ MeV,
 the pasta phase at $T=5$ MeV does not decrease 
when going from $y_p=0.5$ to $y_p=0.3$ for both the TW and DDH$\delta$ models. 
It is also seen that for
symmetric matter the prediction obtained from the  thermodynamical spinodal
is generally quite larger than the values obtained within the CP
calculation, except for the DDH$\delta$ model, when they are similar. This may
  indicate that within a TF calculation larger transition densities would be 
obtained.

Self-consistent calculations at finite temperature, both
  for the pasta phase and for the dynamical spinodal still have to be 
implemented.
These are of particular
  interest because neutrino trapping occurs at finite temperature, and we 
  expect that the interaction of neutrinos with the $npe$ clusters may 
affect the
  neutrino energy deposition in stellar matter. However, while the CP
  calculation fails to predict the transition densities for
  $\beta$-equilibrium matter, we expect that it gives reasonable results for
  the proton fractions of interest for stellar matter with trapped neutrinos,
  $y_p\sim 0.3$.

We show in Table \ref{tabtemp} the highest temperatures for the 
existence of the pasta phase for each of the models discussed in this work 
within the CP approach. We take these values as lower bounds 
for the limit temperature above which the non-homogeneous phase disappears.
It is also worth mentioning that we have searched for the pasta
  structures at temperature steps of 1 MeV,  due
  to the uncertainties mentioned above, and therefore the limiting temperature
  given is not more than an order of magnitude. 
There are two different situations
that had to be considered when the homogeneous phase was taken as the
equilibrium configuration: a) the pasta phase exists but it is not the
preferential state of matter because its free energy comes out higher than
the homogeneous phase and b) the pasta phase does not exist within the
precision of our calculations.

\section{Conclusions}

In the present work we have investigated the extension of the pasta phase for
$npe$ matter described within relativistic density dependent models, namely TW
\cite{tw}, DDH$\delta$ \cite{gaitanos} and GDFM \cite{gogelein}, both at zero
and finite temperatures . The pasta phase was calculated at zero temperature
within Thomas-Fermi and
compared with results obtained in a simplified approach, the coexistence
phases (CP) \cite{pasta1}. Due to the approximate way the surface is described 
within the last approach the pasta phase comes out  smaller with CP than with 
TF: as in  \cite{pasta1} we conclude that a correct description of the surface 
energy and its dependence on the  isospin, temperature and geometry, 
is essential to obtain better results using
the CP formalism. 

The effect of
including the $\delta$-meson was also explicitly investigated: together with
DDH$\delta$ and GDFM we have also considered NL3$\delta$.
It was seen that models with the same description of the isoscalar
  channel and the same symmetry energy at the saturation density, namely the
  couple (NL3, NL3$\delta$), showed 
a smaller non-homogeneous phase for asymmetric matter when the $\delta$ meson
was included. This effect does  not occur for the couple (TW, DDH$\delta$)
because although both have the same description of the isoscalar channel, the
symmetry energy of the DDH$\delta$ at saturation is  smaller. As a result,
the extension of the non-homogeneous phase within the DDH$\delta$ model is 
the largest one for
$\beta$-equilibrium stellar matter and is larger than the corresponding
non-homogeneous phase within the TW model.

Results were compared with 
previous studies done within NLWM and the predictions obtained from the
analysis of the thermodynamical and dynamical spinodals.
One of the main conclusions is that density dependent hadronic models 
generally predict larger
  non-homogeneous phases for asymmetric matter than NLW models. In fact, for
  $\beta$-equilibrium matter a similar conclusion had been taken in
  \cite{camille08} only from the analysis of the crossing of the
  $\beta$-equilibrium EoS with the dynamical spinodal.  Recent
  parametrizations of the 
  Skyrme force, e.g. SLy230a, NRAPR or LNS, showed a similar behavior
  \cite{camille08}. 
We confirm this behavior 
  both within the CP and the TF calculation. 

One important conclusion obtained at $T=0$ MeV is the fact that the transition
density for  $\beta$-equilibrium matter  obtained within a TF calculation
almost coincides with the prediction from the dynamical spinodal. This fact
should be confirmed at finite temperature.  However, for symmetric matter or
for  isospin asymmetries not much smaller that $y_p=0.3$, the TF transition
density is larger than the prediction of the dynamical spinodal. This proton fraction is of
particular interest for neutrino trapped matter for which  $y_p\sim0.3$. In
this case a complete equilibrium calculation should be done.

 The parametrisation GDFM has a very special
behavior with an instability region larger than all the other models,
for quite symmetric matter. However, for very asymmetric matter the
instability region is smaller than the one of other DDH models and is  
 of the order of the  NLW models.

Another important conclusion drawn in the present work is the dependence
  of the pasta phase extension on the isospin asymmetry. For the NLWM it is seen 
clearly that the pasta phase extension decreases
if the isospin asymmetry decreases. For the density dependent models a reduction of the proton 
fraction from 0.5 to 0.3 almost does not affect the pasta phase or may increase it within the CP 
calculation and the dynamical spinodal approach. Within the dynamical spinodal 
approach this behavior is due to the small concavity of the spinodal surface for 
symmetric matter, and the deformation of the spinodal due to the presence of protons, electrons 
and the Coulomb field. A smaller fraction of protons contributes with less repulsion 
and gives a larger instability region. Of course the presence of electrons shields the proton 
repulsion and therefore the effect is not so strong as it would be for charged matter. An adequate 
description of electrons and the Coulomb interaction is important to get a correct 
description of the pasta phase extension. In \cite{maruyama} it was shown 
that the largest pasta phase extension occurs when the inclusion of the 
Coulomb field is done in a self-consistent way.

We have checked which parametrisations fulfill the constraints imposed
in \cite{bao} for the derivative of the symmetry energy and the transition
density and pressure. While the density dependent hadronic models are below 
the lower limit for the
symmetry energy derivative, the NLWM are above the upper
limit. However, both NL3 and NL3$\delta$ together with GDFM fall
within the transition pressure/density limits while TW and DDH$\delta$ have too
large transition densities. The GDFM parametrization
  is the one that satisfies the constraints of
  \cite{bao}  more closely. It seems that the
relation between both quantities, the slope of the symmetry energy and the
transition density, is not, in fact, model free. 

 If we had considered
  the limits on the transition density defined in \cite{link99},
$0.25< P_t< 0.65$ MeV/fm$^3$, all models
  studied here have too small transition pressures except NL3, that lies just
  at the lower border, and TW. 
In \cite{bao}  it was shown that a larger slope $L$
gives rise to a smaller transition density and transition pressure. This
feature is seen when models within the same framework are considered, namely NL3 and
NL3$\delta$. However, no clear trend is seen among the DDH models. In
\cite{camille08} it was shown that the slope of the symmetry energy of the NLW 
models differs from the one of density dependent hadronic models.  
The parametrizations of Skyrme
forces, like SLy230a, NRAPR or LNS, which have values of $L$ at saturation close to the ones of DDH models, also
have the transition densities 
close to the ones of  DDH models and above the limit 0.065 fm$^{-3}$ imposed in
\cite{bao}. This seems to show that a more complete relation between the
transition pressure and transition density and the equilibrium isovector
properties of asymmetric nuclear matter have to be obtained. This could
include constraints on the slope and compressibility of the symmetry energy at
subsaturation densities. For instance, the NLW models have positive
compressibilities of the symmetry energy at subsaturation densities above 0.05
fm$^{-3}$ while the DDH models and the recent Skyrme parametrizations have
negative compressibilities. Another point which should be referred is that the
data obtained from isospin diffusion in heavy-ion reactions correspond to
isospin asymmetries that are far from the ones occurring at
$\beta$-equilibrium matter.  For these large asymmetries we expect that
the contribution from terms beyond the parabolic approximation for the isospin
dependence of the energy density of nuclear matter becomes important \cite{camille08}.

\section*{ACKNOWLEDGMENTS}
This work was partially supported by CNPq and by FCT (Portugal)
under the projects  POCI/FP/81923/2007 and SFRH/BPD/29057/2006.

\end{document}